\documentclass[12pt,journal,draftclsnofoot,a4paper,onecolumn]{IEEEtran}

\usepackage{amssymb,amsthm}
\usepackage{amsfonts,amsmath}
\usepackage{latexsym}
\usepackage[mathscr]{euscript}
\usepackage{graphics}
\usepackage{graphicx}
\usepackage{cite}
\usepackage{cases}

\usepackage[margin=0.82in]{geometry}
\usepackage[T1]{fontenc}%for piotr

\newcommand{\beq}{\begin{equation}}
\newcommand{\eeq}{\end{equation}}

\newcommand{\beqq}{\begin{equation*}}
\newcommand{\eeqq}{\end{equation*}}
\newcommand{\ei}{\end{itemize}}
\newcommand{\bi}{\begin{itemize}}
\newcommand{\ee}{\end{enumerate}}
\newcommand{\be}{\begin{enumerate}}
\newtheorem{definition}{Definition}

\newtheorem{prop}{Proposition}

\newtheorem{lemma}{Lemma}

\theoremstyle{remark}

\newtheorem{remark}{Remark}

\newcommand{\ds}{\displaystyle}

\newcommand{\argmax}[1]{\arg{\hbox{$\underset{#1}{\max}\,$}}}

\newcommand{\ls}[1]
  {\dimen0=\fontdimen6\the\font \lineskip=#1\dimen0
\advance\lineskip.5\fontdimen5\the\font \advance\lineskip-\dimen0
\lineskiplimit=.9\lineskip \baselineskip=\lineskip
\advance\baselineskip\dimen0 \normallineskip\lineskip
\normallineskiplimit\lineskiplimit \normalbaselineskip\baselineskip
\ignorespaces }
\begin{document}

\title{On the Two-user Multi-carrier Joint Channel Selection and Power Control Game}

\author{Majed~Haddad\IEEEauthorrefmark{1},~Piotr~Wiecek\IEEEauthorrefmark{2},~Oussama~Habachi
\IEEEauthorrefmark{3} and~Yezekael~Hayel\IEEEauthorrefmark{1}\\
\IEEEauthorrefmark{1}CERI/LIA, University of Avignon, Avignon, France\\
\IEEEauthorrefmark{2}Faculty of Pure and Applied Mathematics,
Wroclaw University of Technology, Poland\\
\IEEEauthorrefmark{3}XLIM, University of Limoges, Limoges, France\\
%\thanks{Email: \{majed.haddad\}@inria.fr} \\% <-this % stops a space
%\thanks{A preliminary version of this work was presented as a short paper at IFIP Performance 2014 in Turin, Italy \cite{Majed-Performance2014}.}
%\thanks{This research was supported by Grant S40043/K1101 of Wroclaw University of Technology.}
}

\maketitle

\begin{abstract}

In this paper, we propose a hierarchical game approach to model the energy efficiency maximization problem where transmitters \emph{individually} choose their channel assignment and power control. We conduct a thorough analysis of the existence, uniqueness and characterization of the Stackelberg equilibrium. Interestingly, we formally show that a spectrum orthogonalization naturally occurs when users decide sequentially about their transmitting carriers and powers, delivering a binary channel assignment. Both analytical and simulation results are provided for assessing and improving the performances in terms of energy efficiency and spectrum utilization between the simultaneous-move game (with synchronous decision makers), the social welfare (in a centralized manner) and the proposed Stackelberg (hierarchical) game.
For the first time, we provide \emph{tight} closed-form bounds on the spectral efficiency of such a model, including correlation across carriers and users. We show that the spectrum orthogonalization capability induced by the proposed hierarchical game model enables the wireless network to achieve the spectral efficiency improvement while still enjoying a high energy efficiency.
\end{abstract}

\section{Introduction}\label{sec:intro}

Ecological concerns are steadily attracting more and more attention in wireless communications \cite{EE-Commag11,EE_survey_13}. From the operators' perspective, energy efficiency not only has great ecological benefits and represents social responsibility in fighting climate change, but also has significant economic benefits. Therefore, innovative solutions that support traffic increase and maintain a limited energy consumption need to be considered at both system and device levels in order to address environmental and operational costs. Recently, Cisco systems have pointed out that the global mobile data traffic will increase nearly tenfold between 2014 and 2019, giving incentive for service providers to reduce their OpEx by reducing their energy consumption \cite{Cisco-Green-2015}. This suggests to shift from pursuing optimal capacity and spectral efficiency to efficient energy usage when designing wireless networks. Indeed, spectral efficiency has been a traditional requirement of wireless architectures, especially when their access is limited to scarce spectrum. As a result, recent trends in mobile client access tend to support both spectral and energy efficiency at the same time while addressing a wide variety of delay and throughput objectives \cite{AyanogluSurveyEE14}.

\section*{Contributions}
\label{sec:contrib}
\vspace{-0cm}
To address these crucial issues among others, we propose to study energy efficient wireless networks in which we introduce a degree of hierarchy among users. More specifically, we consider energy efficient multi-carrier wireless networks that can be modeled by a decentralized multiple access channel. The network is said to be decentralized in the sense that each user can freely choose his power control policy and carrier assignment in order to selfishly maximize a certain individual performance criterion, called utility (or payoff) in the context of game theoretic studies.

We formally prove that the hierarchical structure of the game \emph{naturally} leads to a spectrum orthogonalization pattern where the components of the network have incentive to transmit on different carriers. This orthogonalization feature across the multiple interfering devices is particularly appealing, not only from an interworking perspective (as a result of reduced infrastructure), but also for increasing both network coverage and data capacity without the need to split the available spectrum. In this sense, we prove that the advantage of the hierarchical (Stackelberg) model that we propose over the simultaneous-move model in \cite{meshkati-jsac-2006} is rather significant.

One could wonder that, as soon as the number of carriers is high, interference can be avoided with high probability. We show next that users still experience interference even when the number of carriers to number of users ratio exceeds a few units, especially for synchronous decision makers. Moreover, to the best of our knowledge, performance bounds have never been derived in the multiple carrier context. This allows us to provide \emph{tight} closed-form bounds on the spectral efficiency of such a model. We formally prove that the spectrum orthogonalization capability induced by the proposed hierarchical game model enables the wireless network to achieve the spectral efficiency improvement while still enjoying a high energy efficiency. In particular, we show that the orthogonalization feature makes correlation over carriers suitable for energy efficient systems as it brings more orthogonalization over the system (and thus leads to higher spectral efficiency), while correlation over users is not suited as it degrades the spectral efficiency.

\section*{Related Literature and Novelty of the Work}
\label{sec:soa}

To reduce the network energy consumption, \cite{GreenCommLetter14} proposed an optimal traffic aware scheme using an online stochastic game theoretic algorithm. In \cite{OFDMA-EE-Jinsong15}, authors proposed a joint transmitter and receiver optimization for the energy efficiency in orthogonal frequency-division multiple-access (OFDMA) systems.
Energy efficient power control game has been first proposed by Goodman \emph{et al.} in \cite{goodman-pcomm-2000} for flat fading channels and re-used by \cite{meshkati-jsac-2006} for multi-carrier systems. \cite{EE12GT} proposed an energy efficient topology control game for wireless ad hoc networks in the presence of selfish nodes.
All these works do not consider hierarchy among different actors in the system. However, as mentioned in \cite{goodman-pcomm-2000} the Nash equilibrium in such games can be very energy inefficient.
Note that the Stackelberg formulation arises naturally in many context of practical interest. For example, the hierarchy is inherent to cognitive radio networks (CRNs) where the user with the higher priority (\emph{i.e.}, the leader or the primary user (PU)) transmits first, then the user with the lower priority (\emph{i.e.}, the follower or the secondary user (SU)) transmits after sensing the spectral environment
\cite{Haykin05,MajedIET08,CR10TMC}. This is also a natural setting for heterogeneous wireless networks due to the absence of coordination among the small cells, and between small cells and macro cells \cite{Femto08Survey,Hoydis2011SmallCells,Majed_wiopt14}. There have been many works on Stackelberg games \cite{bloem-gamecomm-2007,SchaarTWC09,XieInfocom12}, but they do not
consider energy efficiency for the individual utility as defined in
\cite{goodman-pcomm-2000}.
They rather consider transmission rate-type utilities (see
\emph{e.g.,} \cite{ElGamal2008, basar-jota-2002}).

In a prior work \cite{Majed-TVT2014}, we proposed a hierarchical game theoretic model for two-user--two-carrier energy efficient wireless systems. It was shown that, for the vast majority of cases, users choose their transmitting carriers in such a way that if the leader transmits on a given carrier, the follower has incentive to choose the other carrier. One major motivation of this paper is to extend the original problem in \cite{Majed-TVT2014} to some general models that can
be widely used in practice by considering an arbitrary number of carriers.

The work that is most closely related to ours is \cite{samson-twc09}, where the hierarchical game was formulated for the energy efficiency maximization problem in the single carrier system. Notably, it has been proved that, when only one carrier is available for the players, there exists a unique Stackelberg equilibrium.
However, multi-carrier systems have gained intense interest in wireless communications, making the use of multi-carrier transmissions much more appealing for future wireless systems, such as LTE. In fact, the multi-dimensional
nature of such a problem along with the physical properties of the transmission phenomenon make the extension to an arbitrary number of carriers problem much more challenging than the single carrier model. We will see later in the paper that, contrary to \cite{samson-twc09}, we show that, when we come up to study multi-carrier hierarchical games, the degree of freedom increases and \emph{\textbf{leading}} becomes better than \emph{\textbf{following}}. This means that a player can often take advantage of playing first (as the leader), but not always. Indeed, if the players are in the same conditions, a player can improve his utility by playing after observing the action of the other player.

In the light of the above, the paper is structured as follows. The general system model is presented in Sec. \ref{sec:model}. Sec. \ref{sec:game-model} reviews the simultaneous-move game and presents the hierarchical game problem. Then, in Sec. \ref{sec:StackEq}, we characterize the Stackelberg equilibrium, and we evaluate the performance of the Stackelberg approach in Sec. \ref{sec:perf}.
Sec. \ref{sec:simul} provides numerical results to illustrate and validate the
theoretical findings derived in the previous sections. Additional comments and conclusions are provided in Sec. \ref{sec:conc}.

%\vspace{-0.2cm}
\section{Energy Efficient Wireless Network Model}
\label{sec:model}

We consider a wireless network, in which mobile users access to the spectrum in an asynchronous way. We assume that the overall bandwidth can be divided into an arbitrary number of narrow-band carriers $(K \geq 2)$, and that the carriers are narrow enough to undergo flat fading. Let us further suppose that the channels are quasi-static flat fading, i.e., the channel gains are constant during each frame but may change from one frame to the next.

Without the constraint of exclusive
assignment of each carrier for users, we generally formulate the problem of
energy efficiency maximization by allowing that a carrier could be
shared by multiple users. One can think of heterogeneous networks (HetNets) or ultra-dense networks (UDNs) composed of different cellular layers and multiple access technologies. In order to improve the efficiency of spectrum use, multiple overlapping networks operate on the same frequency bands, causing (co-channel) interference, which, in turn, can cause harmful throughput degradation. To be specific, in the following, we will consider a decentralized multiple access channel composed of a leader -- indexed by $1$, having the priority to access the medium, and a follower -- indexed by $2$ that accesses the medium after observing the action of the leader. This setting is particularly relevant for CRNs with the PU as the leader and the SU as the follower, with the difference that no guarantee of service to the PU is considered while sharing the spectrum with the SU. It is also suited for sparse mobile networks in which one may neglect the possibility of simultaneous interference of more than two users. An extension of the proposed model to multiple users with multi-hierarchical levels can be found in \cite{Majed-Networking2015}, where two nearly-optimal algorithms that ensure complete spectrum orthogonalization across users were proposed. Notice that closed-form solutions for the multi-user hierarchical game is in general very difficult to obtain.

Accordingly, for any user $n \in \{1,2\}$ and ${m\neq n}$, the
received signal-to-noise plus interference ratio (SINR) is expressed as
{\small
\beq\label{eq:gamma-mc}
\displaystyle \gamma_{n}^k=\frac{g_{n}^k p_{n}^k}{\sigma^2+ \displaystyle
g_{m}^k p_{m}^k}:=p_n^k \widehat{h}_n^k; \qquad \,\text{for} \,\,\, k=1,\ldots,K
\eeq}
We will call $\widehat{h}_n^k$ the \emph{effective channel gain}, defined as the ratio between the
SINR and the transmission power of the other users over the $k^{th}$ carrier.
$g_n^k$ and $p_{n}^k$ are resp. the fading channel gain and the transmitted power of user $n$ transmitting on carrier $k$, whereas $\sigma^2$ stands for the variance of the Gaussian noise. We statistically
model the channel gains $g_n^k$ to be independent identically distributed (i.i.d.) over the fading coefficients. It follows from the above \mbox{SINR} expression that the strategy chosen by a user affects the performance of other users in the
network through multiple-access interference.

The system model adopted throughout the paper is based
on the seminal paper \cite{goodman-pcomm-2000} that defines the energy efficiency
framework. In order to formulate the power control problem as a game, we first need to define a utility function suitable for
data applications. Let us first define the ``efficiency" function $f(\cdot)$, which
measures the packet success rate. In brief, when \mbox{SINR} is very low, data
transmission results in massive errors and the goodput (rate conditioned to errors) tends to $0$; when
\mbox{SINR} is very high, data transmission becomes error-free and the rate
grows asymptotically to a constant. However, achieving a high \mbox{SINR} level
requires the user terminal to transmit at a high power, which in turn results
in low battery life. This phenomenon is concisely captured by an
increasing, continuous and S-shaped ``efficiency" function $f(\cdot)$. A more detailed discussion of the efficiency function can be found in \cite{meshkati-spmag-2007,BelmegaTSP11,ZapponeTWC13}. The following utility function allows one to measure
the corresponding tradeoff between the transmission benefit
(total goodput over the $K$ carriers) and cost (total power over the $K$ carriers):
%{\scriptsize
\vspace{-0.3cm}
\beq\label{eq:util-mc}
u_n(\mathbf{p_1},\mathbf{p_{2}})=\frac{\displaystyle R_n \cdot \sum_{k=1}^K f(\gamma_{n}^k)}{\displaystyle\sum_{k=1}^K p_{n}^k},
\eeq
%}
where $R_n$ is the transmission
data rate of user $n$ and $\mathbf{p_{n}}$ is the power allocation
vector of user $n$ over all carriers, \emph{i.e.,} $\mathbf{p_{n}}=(p_n^1,\ldots,p_n^K)$. The quantity $R_n$ can be viewed as the gross (transmission) data rate on the radio interface which only depends on the user's application/service induced by high layers such as the transport and the application layers. This target rate may depend on the type of application, but not on the physical layer or the wireless environment of the user. The utility function $u_n$, that has bits per Joule as units, perfectly captures the tradeoff between goodput and battery life, and is
particularly suitable for applications where energy efficiency is
crucial.

\section{The game theoretic formulation}\label{sec:game-model}

One proposal for designing spectrum sharing is through game theory which offers basis to model interactions
between interacting users and develop decentralized and/or distributed algorithms for resource allocation.

\subsection{The simultaneous-move game problem}\label{sec:nash-model}

The interaction between users can be modeled
through a non-cooperative game where each user maximizes his energy efficiency subject to interference constraints, given adversarial decisions. An important solution concept of the game under consideration is the Nash equilibrium, which is a fundamental concept in the strategic games. It is a vector
of strategies (referred to hereafter and interchangeably as
actions), one for each player, $\mathbf{p}^{NE} = \{\mathbf{p_{1}}^{NE},\mathbf{p_{2}}^{NE}\}$ such that no player has incentive to unilaterally change his strategy.

\begin{definition}
A strategy vector $\mathbf{p}^{NE} = \{\mathbf{p_{1}}^{NE},\mathbf{p_{2}}^{NE}\}$ is a Nash Equilibrium (NE) if and only if:
$$
\forall p_1\neq {p_{1}}^{NE},\quad u_1(\mathbf{p_{1}}^{NE},\mathbf{p_{2}}^{NE})\geq u_1(\mathbf{p_{1}},\mathbf{p_{2}}^{NE})
$$
and
$$
\forall p_2\neq {p_{2}}^{NE},\quad u_2(\mathbf{p_{1}}^{NE},\mathbf{p_{2}}^{NE})\geq u_2(\mathbf{p_{1}}^{NE},\mathbf{p_{2}}).
$$
\end{definition}

In what follows, we define a less robust stable strategy vector for non-cooperative games in which the Nash equilibrium is a too strong concept. If there exists an $\epsilon>0$ such that\footnote{The $-n$ subscript on vector $\mathbf{p}$ stands for ``except user $n$".} $(1+\epsilon)u_n(\mathbf{p_{n}}^{\epsilon NE},\mathbf{p}_{-n}^{\epsilon NE})\geq u_n(\mathbf{p_n},\mathbf{p_{-n}}^{\epsilon NE})$ for every action $\mathbf{p_{n}} \neq
\mathbf{p_{n}}^{\epsilon NE}$, we say that the vector $\mathbf{p}^{\epsilon NE} = \{\mathbf{p_{1}}^{\epsilon NE},\mathbf{p_{2}}^{\epsilon NE}\}$
is an $\epsilon$-Nash equilibrium.

The Nash equilibrium concept assumes that the players decide simultaneously. One important framework of non-cooperative games is to assume that one player can observe the decision of the other player before deciding. This concept can be related to asymmetric information/decision in non-cooperative games and is related to the concept of Stackelberg equilibrium.

\subsection{The hierarchical game problem}\label{sec:stack-model}

Hierarchical models in wireless networks are motivated by the idea that the utility
of the leader obtained at the Stackelberg equilibrium can often be improved over his utility obtained at
the Nash equilibrium when the two users play simultaneously
\cite{SchaarTWC09}. It has been proved, in \cite{samson-twc09}, that when only one carrier is available for the players, there exists a unique Stackelberg equilibrium in which both the leader and the follower improve their utilities. The goal is then to find a Stackelberg equilibrium in this bi-level game \cite{bilevel05}.

In this work, we consider a Stackelberg game framework in which, a foresighted follower adapts his power allocation vector $\mathbf{p_{2}}$, based on the
power vector of the leader $\mathbf{p_{1}}$ chosen in advance. The
power allocation of the shortsighted leader will re-embody in
the form of interference introduced to the foresighted follower as given by Eq. (\ref{eq:util-mc}).
At the core lies the idea that, the foresighted follower will extract
the useful asymmetry information in order to make more efficient hierarchical decision making.

\begin{definition} ({\bf Stackelberg equilibrium}):\\
\emph{A vector of actions
$\mathbf{\widetilde{p}}=(\mathbf{\widetilde{p}_1},\mathbf{\widetilde{p}_2})=
(\widetilde{p}_1^{1},\ldots,\widetilde{p}_1^K,\widetilde{p}_2^1,\ldots,\widetilde{p}_2^K)$
is called Stackelberg equilibrium if and only if:
$$
\mathbf{\widetilde{p}_{1}} \in \argmax{\mathbf{p_1}} u_1(\mathbf{p_1},\overline{p}_2(\mathbf{p_1})),
$$
where for all $\mathbf{p_1}$, we have
$$\overline{p}_2(\mathbf{p_1})\in\argmax{\mathbf{p_2}}u_2(\mathbf{p_1},\mathbf{p_2}),
$$
and $\mathbf{\widetilde{p}_2}=\overline{p}_2(\mathbf{\widetilde{p}_1})$.}
\end{definition}

\begin{remark}
Note
that, for sake of
clarity, we will only consider the most interesting (and
non-trivial) regime where the transmit powers are less than
maximal power levels. However, all the results can be easily
extended to the case of finite powers.
\end{remark}

\section{Characterization of the Stackelberg equilibrium}\label{sec:StackEq}

We first determine the best-response function of the follower depending on the action of the leader. This approach is similar to backward induction technique. This result comes directly from Proposition $1$ of \cite{meshkati-jsac-2006}. For making this paper sufficiently self-contained,
we review here the latter proposition.

\begin{prop}[Given in \cite{meshkati-jsac-2006}]\label{prop:follower-power}
Given the power allocation vector $\mathbf{p_1}$ of the leader, the best-response of the follower is
given by
\begin{equation}\label{eq:follower}
\overline{p}_2^k(\mathbf{p_1})= \left\{\begin{array}{lr}\displaystyle
\frac{\gamma^{*}(\sigma^2+g_{1}^k p_{1}^k)}{g_{2}^k},& \mbox{for} \,\,\, k = L_2(\mathbf{p_1})\\
0,& \mbox{for all}\,\,\, k \neq L_2(\mathbf{p_1})
\end{array}
\right.
\end{equation}
with $L_2(\mathbf{p_1})=\argmax{k} \widehat{h}_2^k(p_1^k)$ and $\gamma^*$ is the unique
(positive) solution of the first order equation
\begin{equation}\label{eq:gamma*}
x\,f^{\prime}(x)=f(x)
\end{equation}
\end{prop}
Note that Eq. (\ref{eq:gamma*}) has a unique solution if the efficiency function $f(\cdot)$ is
sigmoidal \cite{rodriguez-globecom-2003}.
The last proposition says that the best-response of the follower is to use only one carrier, the
one such that the effective channel gain is the best.

Let us first present a useful result that will allow us to reduce the complexity of the original problem (with $K$ carriers) to a simpler one where we only focus on the two best carriers.

\begin{prop}
\label{multi:StackNash}
Denote by $B_1$ and $S_1$ two carriers for the leader for which $g_1^k$ is the highest and the second highest respectively, while by $B_2$ and $S_2$ the ones with two highest $g_2^k$ (that is, for the follower). If the Stackelberg game has an equilibrium, then it has an equilibrium where the leader transmits on one of the carriers $\{B_1, S_1\}$, while the follower transmits on one of the carriers $\{B_2, S_2\}$.
\end{prop}

For the clarity of the exposition, all the propositions are proven in the Appendix.% in \cite{tech-report-eff-tcom}.

Given this result, we may only concentrate on strategies where each of the players uses one of his two best carriers. The proposition below gives the algorithm to compute the equilibrium power allocations for both players. Before the proposition, we introduce additional notation, namely
$$\hat{\gamma}=\frac{g^{B_2}_2-g^{S_2}_2}{g^{S_2}_2}.$$
\begin{prop}
\label{multi:power_alloc}
If $B_1\neq B_2$ then equilibrium power allocation of each of the players is
$$\overline{p}_n^k=\left\{ \begin{array}{ll}
\frac{\gamma^*\sigma^2}{g_n^k}&\mbox{when }k=B_n\\
0&\mbox{otherwise}
\end{array}\right.$$
If $B_1=B_2$ then the equilibrium power allocations of the players are computed in three steps:
\begin{enumerate}
\item If $\hat{\gamma}\leq\gamma^*$ then equilibrium power allocation of the leader is
$$\overline{p}_1^k=\left\{ \begin{array}{ll}
\frac{\gamma^*\sigma^2}{g_1^k}&\mbox{when }k=B_1\\
0&\mbox{otherwise}
\end{array}\right.$$
and that of the follower is
$$\overline{p}_2^k=\left\{ \begin{array}{ll}
\frac{\gamma^*\sigma^2}{g_2^k}&\mbox{when }k=S_2\\
0&\mbox{otherwise}
\end{array}\right.$$
Otherwise, go to steps 2 and 3.
\item Find all the solutions $x\leq\frac{\hat{\gamma}}{1+\gamma^*(1+\hat{\gamma})}$ to the equation
\begin{equation}
\label{gamma**}
(x-x^2\gamma^*)f^{\prime}(x)=f(x)
\end{equation}
If there are solutions different than $x=0$, choose the one for which $\frac{f(x)(1-x\gamma^*)}{x}$ is the highest. Let $\beta^{*}$ be this solution.
\item Compare four values\footnote{Of course $V$ can only be computed if $\beta^{*}$ exists.}:
$$V_{B_1}=\frac{f(\beta^{*})(1-\gamma^*\beta^{*})g^{B_1}_1R_1}{\beta^{*}\sigma^2(1+\gamma^*)},\,\,
W_{B_1}=\frac{f(\hat{\gamma})g^{B_1}_1R_1}{\hat{\gamma}\sigma^2},$$
$$U_{S_1}=\frac{f(\gamma^*)g^{S_1}_1R_1}{\gamma^*\sigma^2},\quad
V^0_{B_1}=f^{\prime}(0)\frac{g^{B_1}_1R_1}{\sigma^2(1+\gamma^*)}.$$
If $V_{B_1}$ is the greatest, then equilibrium power allocations of the leader and the follower are
$$\overline{p}_1^k=\left\{ \begin{array}{ll}
\frac{\beta^{*}(1+\gamma^*)\sigma^2}{g_1^k(1-\gamma^*\beta^{*})}&\mbox{when }k=B_1\\
0&\mbox{otherwise}
\end{array}\right.$$
and
$$\overline{p}_2^k=\left\{ \begin{array}{ll}
\frac{\gamma^*(1+\beta^{*})\sigma^2}{g_2^k(1-\gamma^*\beta^{*})}&\mbox{when }k=B_2\\
0&\mbox{otherwise}
\end{array}\right.$$
Next, if $W_{B_1}$ is the greatest, then equilibrium power allocations of the leader and the follower are
$$\overline{p}_1^k=\left\{ \begin{array}{ll}
\frac{\hat{\gamma}\sigma^2}{g_1^k}&\mbox{when }k=B_1\\
0&\mbox{otherwise}
\end{array}\right.$$
and
$$\overline{p}_2^k=\left\{ \begin{array}{ll}
\frac{\gamma^*\sigma^2}{g_2^k}&\mbox{when }k=S_2\\
0&\mbox{otherwise}
\end{array}\right.$$
If $U_{S_1}$ is the greatest, then equilibrium power allocation of the leader is
$$\overline{p}_1^k=\left\{ \begin{array}{ll}
\frac{\gamma^*\sigma^2}{g_1^k}&\mbox{when }k=S_1\\
0&\mbox{otherwise}
\end{array}\right.$$
and that of the follower is
$$\overline{p}_2^k=\left\{ \begin{array}{ll}
\frac{\gamma^*\sigma^2}{g_2^k}&\mbox{when }k=B_2\\
0&\mbox{otherwise}
\end{array}\right.$$
Finally, if $V^0_{B_1}$ is (the only) greatest, then the game has no equilibrium.
\end{enumerate}
\end{prop}

While the formulation of Prop. \ref{multi:power_alloc} is rather complicated, it can be explained in a simpler manner. It describes essentially the way the choice is made by the leader (the follower adjusts to it according to Prop. \ref{prop:follower-power}). If the best carrier of the leader is different than that of the follower, he transmits on his best carrier with power corresponding to \mbox{SINR} $\gamma^*$. If their best carriers are the same, the leader tries to optimize his power on his best carrier $B_1$, by choosing between two powers corresponding to two values of \mbox{SINR}: $\beta^*$, which gives the highest value of the leader's utility if the follower transmits on the same carrier as the leader, creating interference, or $\hat{\gamma}$, which is the smallest value of \mbox{SINR} forcing the follower to change his carrier and reduce the interference on $B_1$. If he can obtain a better utility than the best of the two on some other carrier $S_1$, he chooses to transmit there with the power corresponding to $\gamma^*$.

\begin{remark}
Note that the equilibria computed Prop. \ref{multi:power_alloc} are unique as long as channel gains for different
 carriers are different and as long as $V_{B_1}\neq
 W_{B_1}\neq U_{S_1}$. Also the response of the follower
 at equilibrium is unique as long as channel gains for
 different carriers are different and $W_{B_1}$ is not the
 greatest value in step 3) of the algorithm described by the
 theorem\footnote{Note that, in case there are multiple
 equilibria, because $V_{B_1}=U_{S_1}>W_{B_1}$, the response of the follower to both equilibrium strategies of the leader is unique.}. The matter of uniqueness of the follower's response is obviously very important, as in
 case there are multiple best responses to an equilibrium
 strategy of the leader, the follower has no incentive to
 follow his equilibrium policy. In our case the equilibrium
 strategy can be imposed to the follower when he has multiple
 best responses to the leader's policy by using a simple
 trick: whenever $W_{B_1}$ appears to be the greatest in step
 3), the leader has to use power infinitesimally smaller than
 that prescribed by his equilibrium policy. This gives him a
 minimally smaller utility, but at the same time makes the
 best response of the follower unique.
\end{remark}

\begin{remark}
The reasoning behind Prop. \ref{multi:StackNash} works also for the model where powers that players can use are limited to the sets $[0,P_{max}]$, so also in this case each player transmits on only one of his two best carriers.  Prop. \ref{multi:power_alloc} gives the form of a Stackelberg equilibrium in case each user has enough power in $[0,P_{max}]$ to reach the $\mbox{SINR}$ $\gamma^*$. Otherwise, it can be shown that all the computations of Prop. \ref{multi:power_alloc} can be repeated under assumption that whenever the desired value of the $\mbox{SINR}$ cannot be reached within the constrained regime, the users transmit at their maximum power. In that way we also obtain an equilibrium in the model. However, considering power constraints will induce additional cases where the equilibrium is such that some users transmit with their maximum power $P_{max}$, complicating the formulation of the results, without changing their general sense.
\end{remark}

The next proposition characterizes the degenerate case when there is no equilibrium in the Stackelberg game.
\begin{prop}
\label{multi:eps_equilibria}
The Stackelberg game has no equilibrium iff $B_1=B_2$, $\hat{\gamma}>\gamma^*$ and
\begin{equation}
\begin{array}[t]{l}
\label{eps_cond}
\hspace{-0.5cm}f^{\prime}(0)>
\max\left\{\frac{f(\hat{\gamma})(1+\gamma^*)}{\hat{\gamma}},\frac{f(\gamma^*)(1+\gamma^*)}{\gamma^*}\frac{g_1^{S_1}}{g_1^{B_1}},\frac{f(\beta^{*})(1-\gamma^*\beta^{*})}{\beta^{*}} \right\},
\end{array}
\end{equation}
but for any $\epsilon>0$ there are $\epsilon$-equilibria of the form
$$\overline{p}_1^k(\epsilon)=\left\{ \begin{array}{ll}
\alpha({\epsilon})&\mbox{when }k=B_1\\
0&\mbox{otherwise}
\end{array}\right.$$
for the leader and
$$\overline{p}_2^k(\epsilon)=\left\{ \begin{array}{ll}
\frac{\gamma^*(\sigma^2+g_1^k\alpha(\epsilon))}{g_2^k}&\mbox{when }k=B_2\\
0&\mbox{otherwise}
\end{array}\right.$$
for the follower, where $\alpha(\epsilon)$ is an arbitrarily small value, guaranteeing that the utility of the leader is within $\epsilon$ from $V_{B_1}^0$. \end{prop}

\begin{remark}
It is important to notice that the case considered in Proposition \ref{multi:eps_equilibria} is indeed possible for some sigmoidal function $f$. One example of such $f$ is one of the form:
$$f(x)=\left\{ \begin{array}{ll}
\frac{1}{\sqrt{1-x}}-1;&x\leq\frac{3}{4}\\
\frac{7+\sqrt{17}}{4}-\frac{13+3\sqrt{17}}{32x+2\sqrt{17}-18};&x\geq\frac{3}{4}
\end{array}\right.$$
One can check that $f$ is concave on interval $[0,\frac{3}{4}]$ and convex on $[\frac{3}{4},\infty)$. Moreover, $f$ and $f^{\prime}$ are continuous and $\lim_{x\rightarrow\infty}f(x)=\frac{7+\sqrt{17}}{4}$, so it is definitely a sigmoidal function. It is straightforward to compute that $\gamma^*=1$ for this function. Unfortunately, Equation (\ref{gamma**}) has no solutions on $(0,\infty)$, which can be computed either numerically or using Taylor expansion of the function $\sqrt{1-x}$. Finally, $f^{\prime}(0)=\frac{1}{2}$, and so for $g^{B_2}_2\gg g^{S_2}_2$ and $g^{B_1}_1\gg g^{S_1}_1$, the inequality (\ref{eps_cond}) will be satisfied.
\end{remark}

On the other hand, any of the two following assumptions:\\
{\bf (A1)} $f^{\prime}(0^+)=0$,\\
{\bf (A2)} $f^{\prime}(0^+)>0$ and $\frac{f^{\prime\prime}(0^+)}{f^{\prime(0^+)}}>2\gamma^*$,\\
implies that (\ref{eps_cond}) is never satisfied, and so the game under consideration always has an equilibrium. In particular, for the most standard form of $f$ \cite{meshkati-jsac-2006},
$$f(x)=(1- e^{-x})^M, \qquad M>1$$
not only there always exists an equilibrium in the Stackelberg model (because $f$ satisfies (A1)), but also the procedure in Proposition \ref{multi:power_alloc} slightly simplifies, as:

\begin{enumerate}
\item Eq. (\ref{eq:gamma*}) can be written as $Mx=e^x-1$,
\item Eq. (\ref{gamma**}) can be written as $M(x-x^2\gamma^*)=e^x-1$, moreover it has exactly one positive solution.
\end{enumerate}

\section{Performance Evaluation}\label{sec:perf}

This section is dedicated to present some key properties and performances of the Stackelberg equilibrium we derived in the previous section. We first study the individual performance of each player. Then, we evaluate the global performance of the system in terms of energy efficiency and spectral efficiency.

\subsection{Individual Performance Evaluation}

\subsubsection{Spectrum orthogonalization} \label{sec:coord}

In this section, we shall first look for what values of channel gains for each of the users there is a possibility that both the leader and the follower transmit on the same carrier. In the sequel, we will refer to the case where users transmit on the same carrier as there is no orthogonalization between users, i.e., $\exists \, k \mid \overline{p}_n^k\neq0$ for $n=\{1,2\}$. Then, we will compute the probability that there is no orthogonalization between the players.

\begin{prop}
\label{multi:no_coordination_region}
The set of $\{g_n^1,\ldots,g_n^K\}$, $n=1,2$ for which there is no orthogonalization between users is a proper subset of the set $G_0$ of $g_n^k$s satisfying
\begin{equation}
\label{chfol:multi}
B_1=B_2\mbox{ and }
g^{B_n}_n\geq (1+\gamma^*)g^{S_n}_n; \,\,\mbox{for }n=1,2.
\end{equation}
\end{prop}
Note that $G_0$ is exactly the set of $g_1^k$s for which there is no orthogonalization in the simultaneous-move game considered in \cite{meshkati-jsac-2006}. Thus, introducing hierarchy in the game induces more spectrum orthogonalization than there was in the simultaneous-move scenario.

In the next proposition, we will show that the probability of no orthogonalization between the players is always small and decreases fast as the number of carriers grows.

\begin{prop}
Assume that the channel gains for different carriers of each of the users are i.i.d. Rayleigh random variables. Then, the probability that there is no orthogonalization between the players at the Stackelberg equilibrium is bounded above by
\begin{multline}
\label{eq:probab_nocord}
(1+\gamma^*)\mathcal{B}(1+\gamma^*,K)\left[ \frac{K-1}{K}+(1+\gamma^*)\mathcal{B}(1+\gamma^*,K)\right]
\sim \mathcal{O}(K^{-(1+\gamma^*)})
\end{multline}
where $\mathcal{B}$ denotes the Beta function, which is the exact probability of no orthogonalization in the simultaneous-move version of the model.
\label{multi:no_coordination_prob}
\end{prop}

\begin{remark}
In the above proposition, we suppose that the channel gains of different players are not correlated, which is typically the case when carriers are far enough \cite{proakis01}. Otherwise, the probability computed there can be treated as an upper bound for respective probabilities, when there is a positive correlation between different carriers of each of the users, which is much more realistic. We will see later in the paper (see Fig. \ref{fig:ncordn-corr} and \ref{fig:ncords-corr}) that, in the case of positive correlation over carriers, these probabilities will be even smaller (and so faster decreasing to $0$).
\end{remark}

\begin{remark}
Now, the opposite situation to that analyzed in Prop. \ref{multi:no_coordination_prob} is when both users experience the same channel gains. The probability that there is no orthogonalization between the players in the Stackelberg game is then bounded above by
\beq\label{eq:probab_nocord1}
K(1+\gamma^*)\mathcal{B}(1+\gamma^*,K)
\eeq
which is still decreasing to $0$ as $K$ goes to infinity, but $K$ times bigger than the bound in Eq. (\ref{eq:probab_nocord}).
The intuition behind this is that, if the channels of
different users are not correlated, then with probability $(K-1)/K$ users have different best channels and with only $1/K$ users have the same
best channels (and interference is an issue in this case). If users experience the same channel gains, they have the same best channel with probability $1$. Also, if the number of carriers $K$ is big, both users will have two best carriers of similar quality as the channel gains are chosen at random, so the probability that they choose the same carrier becomes very small (see Fig. \ref{fig:ncordn} and \ref{fig:ncords}).
\end{remark}

\subsubsection{Payoffs comparison} \label{sec:payoff}
The leader is not worse off on introducing hierarchy (which is always the case in Stackelberg games if both the leader and the follower use their equilibrium policies), but the follower loses on it in some cases. The proposition below gives more insights on what the latter depends on.

\begin{prop}
\label{prop:payoffs}
For any sigmoidal function $f$ the following three situations are possible:
\begin{enumerate}
\item $B_1\neq B_2$. Then, for both players, the payoff in the Stackelberg game is the same as that in the simultaneous-move game.
\item Both players use the same carrier $B_1=B_2$ in equilibria (or $\epsilon$-equilibria) of simultaneous-move and Stackelberg games. Then, the payoff of the follower in the Stackelberg game is always bigger than what he receives in the simultaneous-move game.
\item $B_1=B_2$ and both players use different carriers in equilibria of simultaneous-move and Stackelberg games: the leader in the Stackelberg game uses $B_1$, but in the simultaneous-move game he uses $S_1$ in equilibrium, the follower in the Stackelberg game uses $S_2$, while in the simultaneous-move game he uses $B_2$ in equilibrium. Then, the payoff of the follower in the Stackelberg game is smaller than what he receives in simultaneous-move game.
\end{enumerate}
\end{prop}

\subsubsection{Comparison between leading and following} \label{sec:lead-foll}

It is known from \cite{samson-twc09}, that if there is only one carrier available for the players, it is always better to be the follower than to be the leader. The situation changes when the number of carriers increases.
\begin{prop}
\label{prop:lead_vs_follow}
Suppose that the Stackelberg game has exact equilibria both when player 1 is the leader and when he is the follower.
Then, the utility at Stackelberg equilibrium of player $1$ if he is the leader is not less than his utility if he is the follower if one of the following conditions is satisfied:
\begin{enumerate}
\scriptsize
\item $B_1\neq B_2$.
\item $B_1=B_2$ and $\min\{\frac{g_1^{B_1}}{g_1^{S_1}},\frac{g_2^{B_2}}{g_2^{S_2}}\}\leq 1+\gamma^*$%\vspace{0.2cm}
\item $B_1=B_2$ and for $i=1,2$, $j\neq i$, $$\frac{f\left(\frac{g_i^{B_i}}{g_i^{S_i}}-1\right)}{\frac{g_i^{B_i}}{g_i^{S_i}}-1}\geq \max\left\{\frac{f(\gamma^*)g_j^{S_j}}{\gamma^*g_j^{B_j}},\frac{f(\beta^{*})(1-\gamma^*\beta^{*})}{\beta^{*}(1+\gamma^*)}\right\}$$
\item $B_1=B_2$, $$\frac{f(\beta^{*})(1-\gamma^*\beta^{*})}{\beta^{*}(1+\gamma^*)}\geq \max\left\{\frac{f(\gamma^*)g_1^{S_1}}{\gamma^*g_1^{B_1}},\frac{f\left(\frac{g_2^{B_2}}{g_2^{S_2}}-1\right)}{\frac{g_2^{B_2}}{g_2^{S_2}}-1}\right\}$$ and $$\frac{f\left(\frac{g_1^{B_1}}{g_1^{S_1}}-1\right)}{\frac{g_1^{B_1}}{g_1^{S_1}}-1}\geq \max\left\{\frac{f(\gamma^*)g_2^{S_2}}{\gamma^*g_2^{B_2}},\frac{f(\beta^{*})(1-\gamma^*\beta^{*})}{\beta^{*}(1+\gamma^*)}\right\}$$
\item $B_1=B_2$, $$\frac{f(\gamma^*)g_1^{S_1}}{\gamma^*g_1^{B_1}}\geq \max\left\{\frac{f\left(\frac{g_2^{B_2}}{g_2^{S_2}}-1\right)}{\frac{g_2^{B_2}}{g_2^{S_2}}-1},\frac{f(\beta^{*})(1-\gamma^*\beta^{*})}{\beta^{*}(1+\gamma^*)}\right\}$$ $$\frac{f(\beta^{*})(1-\gamma^*\beta^{*})}{\beta^{*}(1+\gamma^*)}\geq \max\left\{\frac{f(\gamma^*)g_2^{S_2}}{\gamma^*g_2^{B_2}},\frac{f\left(\frac{g_1^{B_1}}{g_1^{S_1}}-1\right)}{\frac{g_1^{B_1}}{g_1^{S_1}}-1}\right\}$$ and $$\frac{g_1^{B_1}}{g_1^{S_1}}\leq\frac{1+\beta^{*}}{1-\gamma^*\beta^{*}}$$
\item $B_1=B_2$, $$\frac{f\left(\frac{g_2^{B_2}}{g_2^{S_2}}-1\right)}{\frac{g_2^{B_2}}{g_2^{S_2}}-1}\geq \max\left\{\frac{f(\gamma^*)g_1^{S_1}}{\gamma^*g_1^{B_1}},\frac{f(\gamma^*)(1-\gamma^*\beta^{*})}{\gamma^*(1+\gamma^*)}\right\}$$ and $$\frac{f(\beta^{*})(1-\gamma^*\beta^{*})}{\beta^{*}(1+\gamma^*)}\geq \max\left\{\frac{f(\gamma^*)g_2^{S_2}}{\gamma^*g_2^{B_2}},\frac{f\left(\frac{g_1^{B_1}}{g_1^{S_1}}-1\right)}{\frac{g_1^{B_1}}{g_1^{S_1}}-1}\right\}$$
\end{enumerate}
\end{prop}

Although the formulation of the proposition is rather complicated, its general meaning is simple. It states that, in most of the cases, different users have different best carriers, so there is no difference between leading and following. The two remaining cases are when both players have the same best carrier. In the first one, each of the players has only one good carrier and the same for both. This situation reduces to the problem considered in \cite{samson-twc09} where only one carrier is available, and so every user can obtain better energy efficient utility by decreasing its priority from leading to following. The reason behind this phenomenon is basically the construction of the energy-efficient utility. In the simultaneous-move version of this model each user transmits with the power corresponding to the \mbox{SINR} $\gamma^*$. Under Stackelberg regime, the leader can increase his utility by reducing his power consumption to the level corresponding to the \mbox{SINR} $\beta^*<\gamma^*$, which reduces the overconsumption due to interference. The optimal answer of the follower will still be to use the power giving him the \mbox{SINR} of $\gamma^*$ though. The result of the shift in the power used by the leader without a similar change in that of the follower is that both utilities increase simultaneously, but the increase of the utility of the follower is bigger than that of the leader. In the second case, both players have the same best carrier but one of them prefers to use his second best carrier instead (that is -- the second best carrier is not much worse than the best one). In this situation, it is the leader who is better off on introducing hierarchy, so this becomes similar to most of the Stackelberg models\footnote{Notice that in basic standard economic problems, it is commonly known that a firm does always better by preempting the market and setting its output level first (e.g., in Cournot-like competition games) \cite{Tirole91}.
}. It is worth noting though that if $g_n^k$ are i.i.d. Rayleigh random variables, one of the two first cases of Proposition \ref{prop:lead_vs_follow} will occur with probability significantly bigger than $1-(1+\gamma^*)\mathcal{B}(1+\gamma^*,K)\left[ \frac{K-1}{K}+(1+\gamma^*)\mathcal{B}(1+\gamma^*,K)\right]$, and so it will be very close to $1$ even for small values of $K$. We will show later in the paper (see Figure \ref{fig:ee_K_user}) that, in practice, it is the last situation that prevails whenever the players have at least two carriers at their disposal.

\subsection{System Performance Evaluation}
\subsubsection{Energy efficiency} \label{sec:soc}

Let us now compute the social welfare in our model, defined as the sum of utilities of both players.
In the following proposition, we give upper bounds on the possible decrease of social welfare when we introduce hierarchy in the game, as well as a bound on the ratio of the maximum social welfare obtainable and that of Stackelberg equilibrium in the game. The latter can be treated as the price of anarchy \cite{Moustakas_PoA08} in our game.

\begin{prop}
\label{prop:soc_welf}
The social welfare when the players apply Stackelberg equilibrium policies equals both maximum social welfare obtainable in the game and social welfare in Nash equilibrium of the simultaneous-move game whenever $B_1\neq B_2$. When $B_1=B_2$, the social welfare in Stackelberg equilibrium:
\begin{enumerate}
\item Is at most $\frac{R_1g_1^{B_1}+R_2g_2^{B_2}}{R_1g_1^{B_1}+R_2g_2^{S_2}}$ times worse than that in simultaneous-move game equilibrium.
\item Is at most $\frac{R_1g_1^{B_1}+R_2g_2^{B_2}}{R_1g_1^{S_1}+R_2g_2^{S_2}}$ times worse than the maximum social welfare obtainable in the game.
\end{enumerate}
\end{prop}
Note that, when $g_n^k$ are i.i.d. Rayleigh variables (as assumed in Proposition \ref{multi:no_coordination_prob}), then a) the region where Stackelberg equilibrium is not the social optimum shrinks fast as the number of carriers increases; b) even in case there is no orthogonalization in Stackelberg equilibrium, the ratios appearing in the above proposition are small with probability increasing with the number of carriers.

\subsubsection{Spectral efficiency} \label{sec:SE}

Along with energy efficiency, spectral efficiency -- defined as the throughput per unit of bandwidth -- is one of the key performance evaluation criteria for wireless network design. These two conflicting criteria can be linked through their tradeoff \cite{Heliot_SE_EE_tcom12,Joung_SE_EE_jsac13}. Therefore, it is often imperative to make a tradeoff
between energy efficiency and spectral efficiency. In the following, we give a closed-form expression of the lower bound on the sum spectral efficiency of the proposed Stackelberg model.

\begin{prop}
\label{prop:spec_ef}
{\normalsize
The spectral efficiency in case there is a orthogonalization between the users at the Stackelberg equilibrium is strictly bigger than}
%\begin{eqnarray}
%\begin{dmath}
\begin{multline}\label{eq:se}
%\hspace*{-1.7cm}
\ds\log_2(1+\gamma^*)\bigg[ 1-(1+\gamma^*)\mathcal{B}(1+\gamma^*,K)\cdot
\left.\left( \frac{K-1}{K}+(1+\gamma^*)\mathcal{B}(1+\gamma^*,K)\right)\right]
%\end{dmath}
\end{multline}
%\end{eqnarray}
which is equal to the spectral efficiency in the simultaneous-move game.
\end{prop}

The computation done in Proposition \ref{prop:spec_ef} holds in case there is orthogonalization between the players. This means that this is only a lower bound for the total spectral efficiency in our model. However, by Proposition \ref{multi:no_coordination_prob}, it becomes very tight as $K$ goes to infinity. An easy consequence of this is that the spectral efficiency in the limit model (with an infinite number of carriers) can be computed exactly, and is equal to $\log_2(1+\gamma^*)$.
Notice that, when users experience the same Rayleigh channel gains, the spectral efficiency in case there is a orthogonalization between the users at the Stackelberg equilibrium is strictly bigger than
\beq\label{eq:se1}
\log_2(1+\gamma^*)\left[ 1-K(1+\gamma^*)\mathcal{B}(1+\gamma^*,K)\right]
\eeq

\section{Simulation Results}\label{sec:simul}

\begin{figure}[t]
\centering
%\vspace*{-0.8cm}
\hspace*{-0.5cm}
\includegraphics[height = 4.4cm,width=9.5cm]{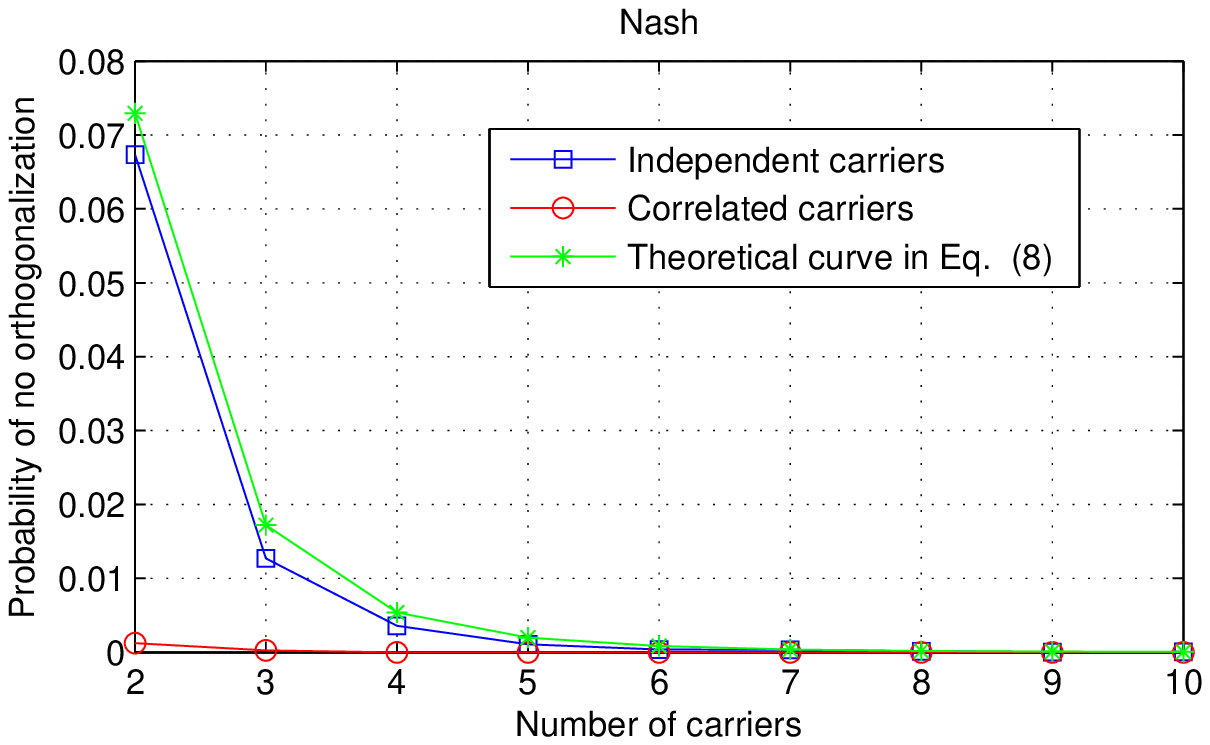}
%\vspace*{-0.5cm}
\caption{The probability of no orthogonalization between the players at the Nash equilibrium with correlation over carriers.}
\label{fig:ncordn-corr}
%\end{figure}
%\begin{figure}[t]
\centering
\vspace*{0.5cm}
\hspace*{-0.5cm}
\includegraphics[height = 4.5cm,width=9.5cm]{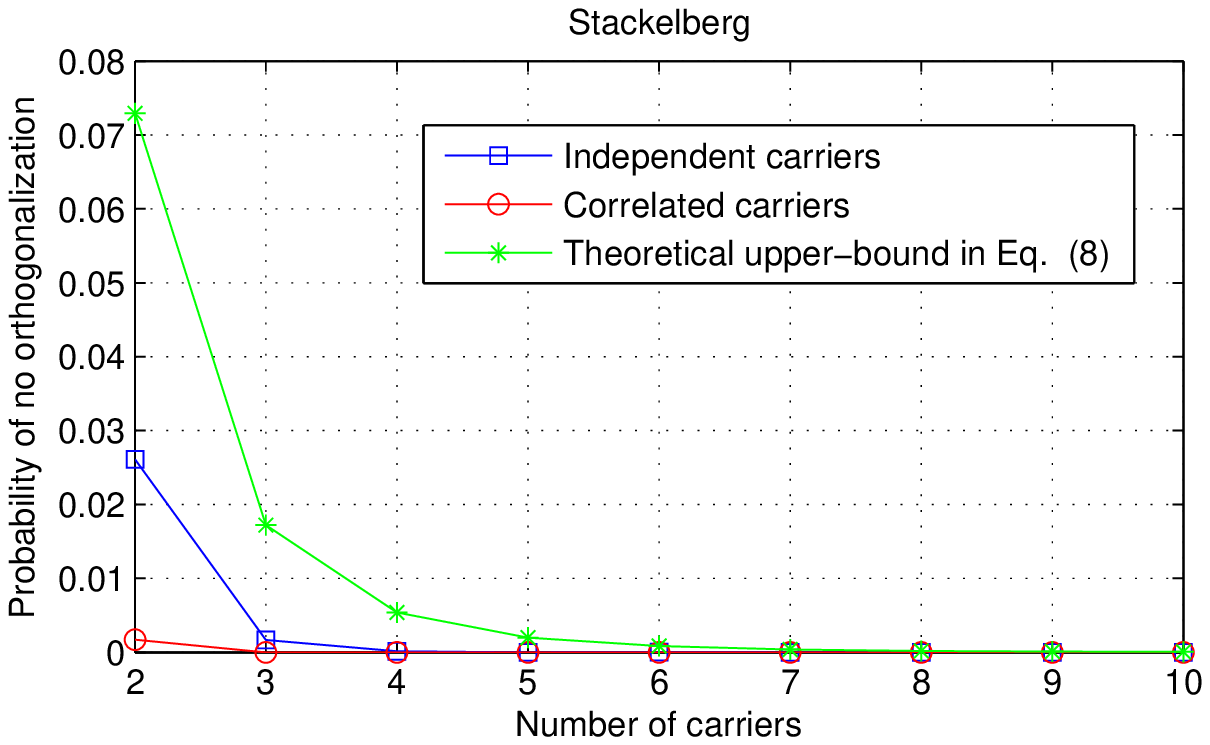}
%\vspace*{-0.3cm}
\caption{The probability of no orthogonalization between the players at the Stackelberg equilibrium with correlation over carriers.}
\label{fig:ncords-corr}
\end{figure}

We consider the energy efficiency function, $f(x)=(1-e^{-x})^M$, well-known in power allocation games, where $M=100$ is the block length in bits. For this efficiency function, $\gamma^*\simeq 6.4$ (or $8.1$ dB). Simulations were carried out using a rate $R_n =1$ bps for $n=\{1,2\}$. We have simulated $10000$ scenarios to remove the random effects from Rayleigh fading.

\subsection{The probability of no orthogonalization}

Let us first consider a quasi-static correlated Rayleigh-fading channel model. Fig. \ref{fig:ncordn-corr} and \ref{fig:ncords-corr} reflect the effect of the correlation over {\it{\textbf{carriers}}} (i.e., the correlation between different carriers of each of the users) on the probability of no orthogonalization for the simultaneous (Nash) and the hierarchical (Stackelberg) game respectively. The correlation model follows the model in \cite{Corr_Rayleigh05}.
As we expected in Section \ref{sec:coord} (see Remark 4), results show that, in the case of correlated carriers, the probability of no orthogonalization is smaller and so faster decreasing to $0$ even for a moderate number of carriers $K$.

From now on, we will only consider the case of no correlation between different carriers of each of the users. In case of correlated carriers, performance results obtained in the remainder have to be considered as a worst case performance.

\begin{figure}[t]
\centering
%\vspace*{-.5cm}
\hspace*{-0.5cm}
\includegraphics[height = 4.4cm,width=9.5cm]{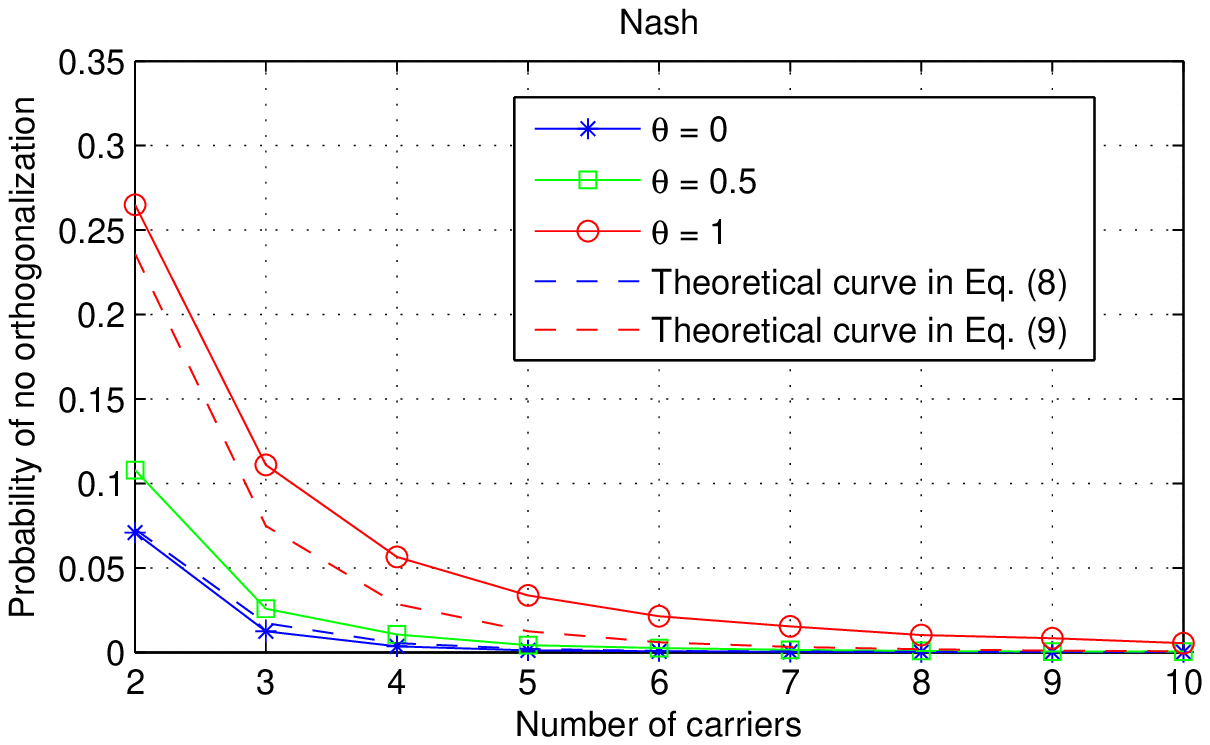}
\vspace*{-0.6cm}
\caption{The probability of no orthogonalization between the players at the Nash equilibrium with correlation over users.}
\label{fig:ncordn}
%\end{figure}
%\begin{figure}[t]
\centering
\vspace*{0.5cm}
\hspace*{-0.4cm}
\includegraphics[height = 4.4cm,width=9.5cm]{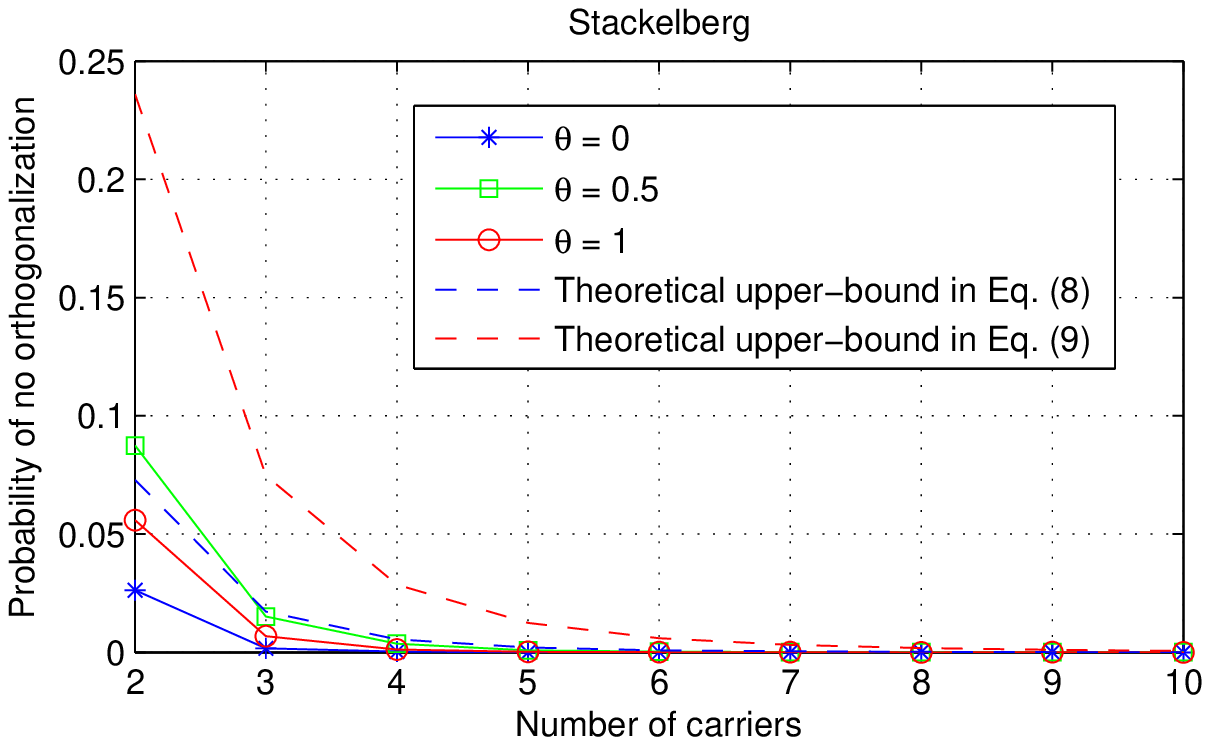}
%\vspace*{-0.8cm}
\caption{The probability of no orthogonalization between the players at the Stackelberg equilibrium with correlation over users.}
\label{fig:ncords}
\end{figure}

Fig. \ref{fig:ncordn} and \ref{fig:ncords} investigate the effect of the correlation over {\it{\textbf{users}}} (i.e., the correlation between different users' fading channel) on the probability of no orthogonalization for the Nash and the Stackelberg game respectively. The correlation factor modeling the dependencies between the users is $\theta$. In both figures, results show that, as the correlation between different users decreases, the probability of no orthogonalization gets even smaller and so faster decreasing to $0$, which corresponds to what Remark 5 claims. In order to assess the accuracy of the theoretical bounds, we also compare the simulated probability of no orthogonalization with the theoretical upper-bounds. More specifically, for i.i.d. users, we compare theoretical curve derived in Eq. (\ref{eq:probab_nocord}) with simulated curve for $\theta = 0$. For correlated users, we compare theoretical curve in Eq. (\ref{eq:probab_nocord1}) with simulated curve for $\theta = 1$. We see that the simulated and theoretical curves match pretty well. Now, when we look at the Stackelberg equilibrium in Fig. \ref{fig:ncords}, it is clearly illustrated that the theoretical upper-bounds derived in Eq. (\ref{eq:probab_nocord}) and Eq. (\ref{eq:probab_nocord1}) turn out to be greater than the simulated probabilities of no orthogonalization, which confirms the accuracy of the results. Remember that the theoretical curves derived in Eq. (\ref{eq:probab_nocord}) and Eq. (\ref{eq:probab_nocord1}) correspond to the exact probability of no orthogonalization in the simultaneous-move game, but are only upper-bounds in the hierarchical version of the model, which is clearly confirmed by Fig. \ref{fig:ncordn} and \ref{fig:ncords}.
\begin{figure}[t]
\centering
\hspace*{-0.5cm}
\includegraphics[height = 4.4cm,width=9.5cm]{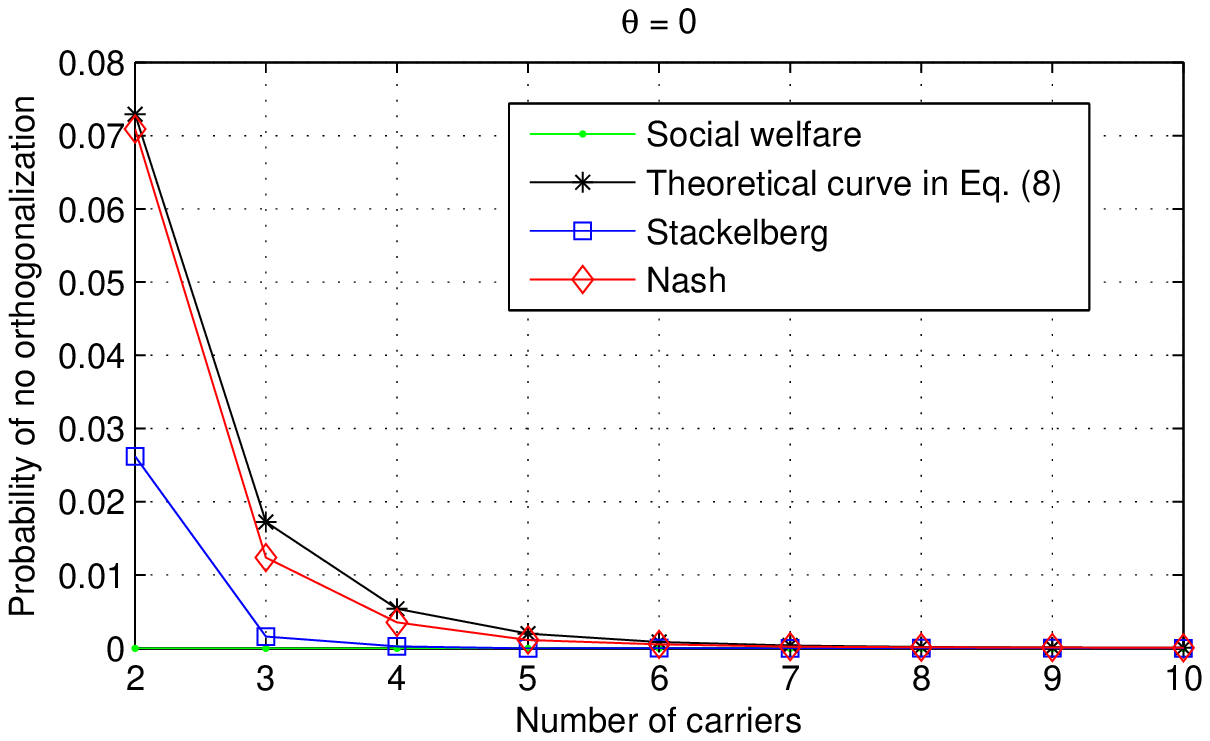}
%\vspace*{-5.8cm}
\caption{The probability of no orthogonalization between the players as a function of the number of carriers with independent users.}
\label{fig:ncord0}
%\end{figure}
%\begin{figure}[t]
\centering
\vspace*{0.5cm}
\hspace*{-0.5cm}
\includegraphics[height = 4.4cm,width=9.5cm]{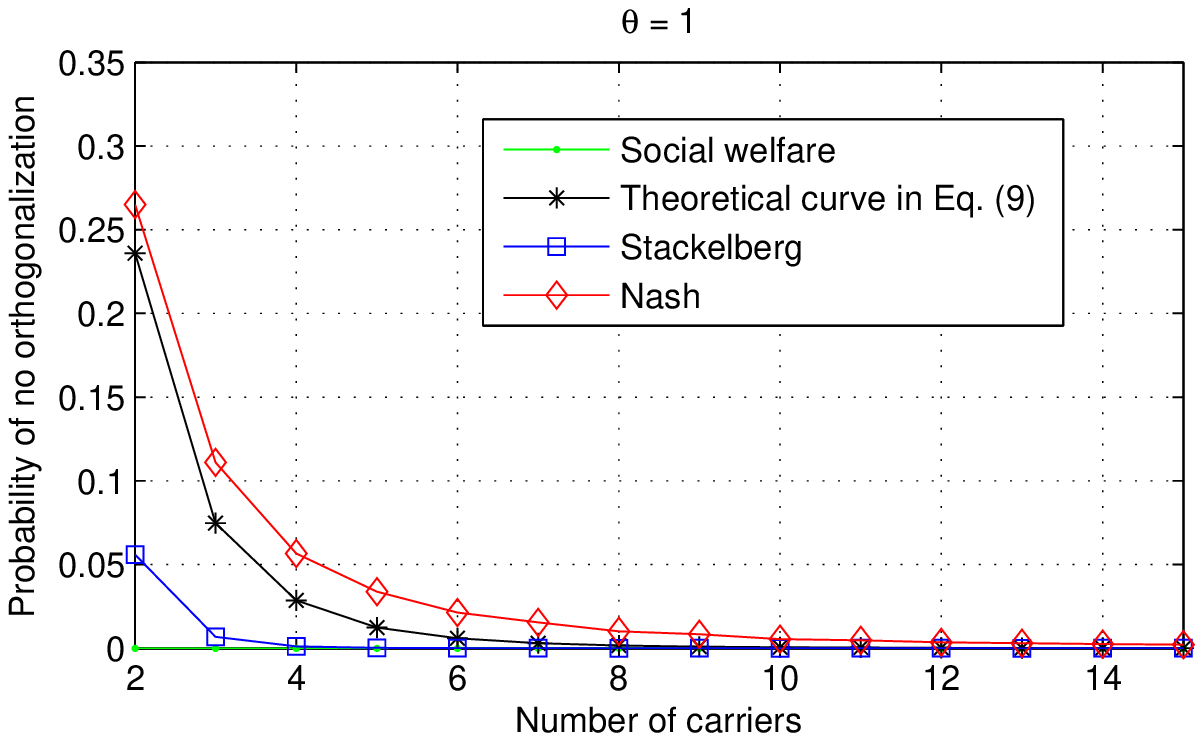}
%\vspace*{-5.8cm}
\caption{The probability of no orthogonalization between the players as a function of the number of carriers with correlated users.}
\label{fig:ncord1}
\end{figure}

Fig. \ref{fig:ncord0} and \ref{fig:ncord1} depict the probability of no orthogonalization for different schemes considering independent users (\emph{i.e.,} for $\theta = 0$) and correlated users (\emph{i.e.,} for $\theta = 1$) respectively. Both curves follow the same trend, tending to increase the orthogonalization between the users as the number of carriers grows, which validates the obtained theoretical results. A rather significant gap between Nash and Stackelberg curves suggests that introducing hierarchy results in much more orthogonalization between the players. Particularly noteworthy is the fact that, at the social optimum, we always obtain strict orthogonalization between users. This means that, in a centralized system, if maximizing the energy efficiency is the goal, introducing hierarchy moves the solution closer to the social optimum.

To sum it up, we can argue that correlation across carriers is a suitable feature as it brings more orthogonalization (and thus leads to a better spectral efficiency), desirable from the social point of view, while correlation across users is not suited as it increases the probability of no orthogonalization. This results is of practical interest as it suggests that designing the power control for multi-carrier networks shall be developed tailored to the physical properties of the transmission phenomenon.
\begin{figure}[t]
\centering
%\vspace*{-5.6cm}
\hspace*{-0.2cm}
\includegraphics[height = 4.4cm,width=8.5cm]{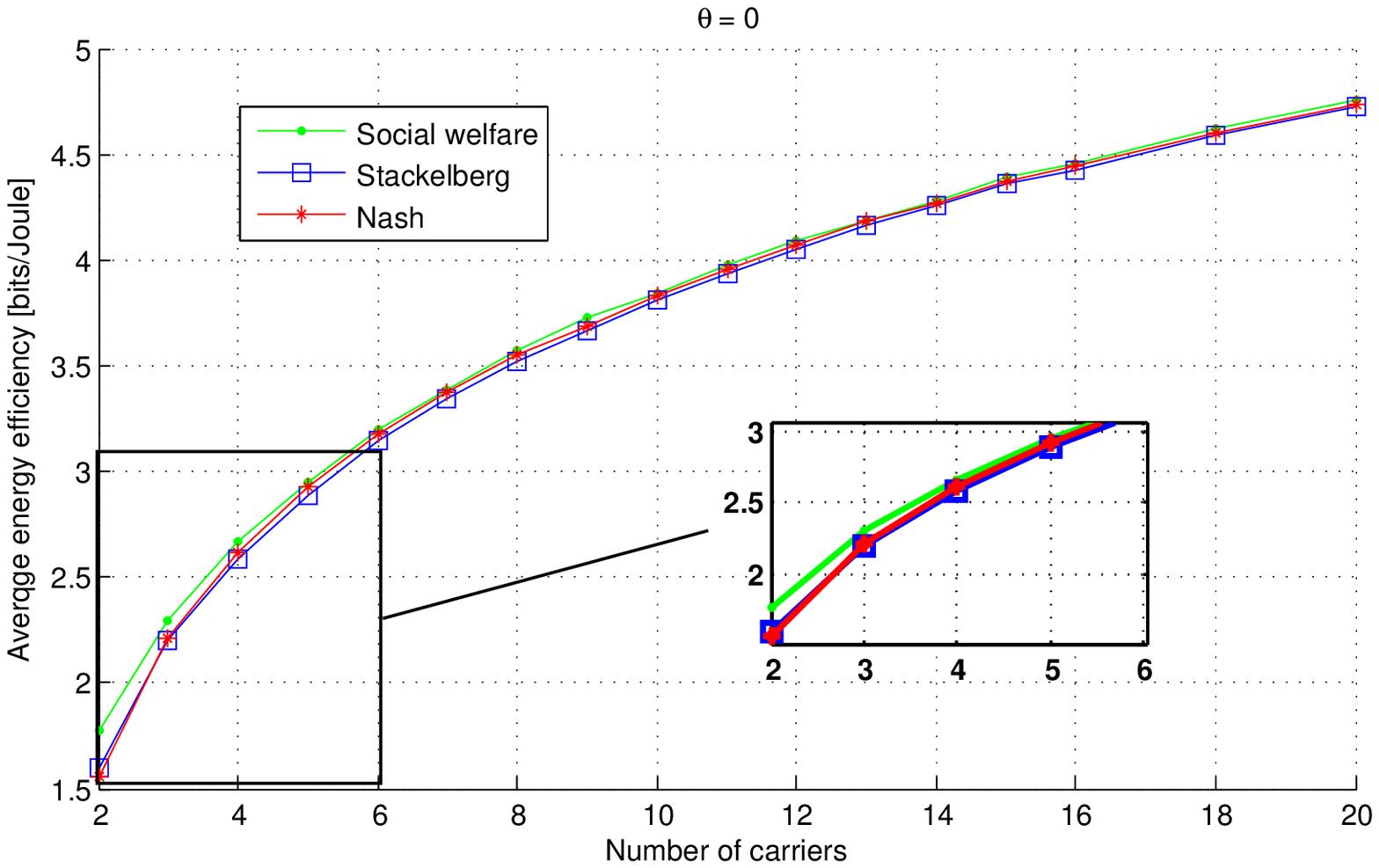}
%\vspace*{-4.3cm}
\caption{Average energy efficiency with independent users.}
\label{fig:ee_K_theta0}
%\end{figure}
%\begin{figure}[t]
\centering
\vspace*{0.5cm}
\hspace*{-0.5cm}
\includegraphics[height = 4.4cm,width=9.5cm]{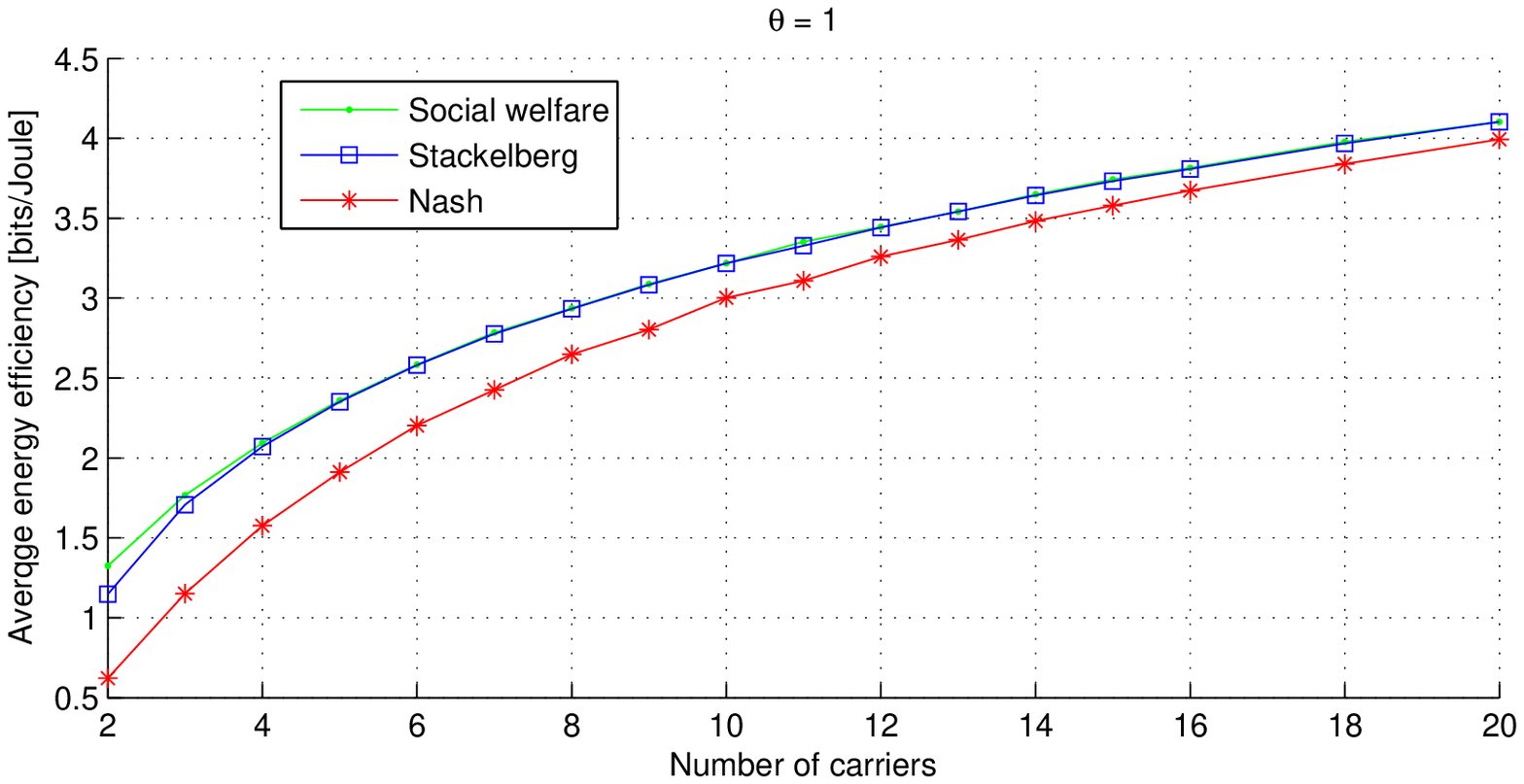}
%\vspace*{-5.8cm}
\caption{Average energy efficiency with correlated users.}
\label{fig:ee_K_theta1}
\end{figure}

\subsection{Energy efficiency}

We then resort to plot the average energy efficiency at the equilibrium for increasing number of carriers $K$. The curves obtained in Fig.~\ref{fig:ee_K_theta0} for independent users (\emph{i.e.,} for $\theta = 0$) exhibit a different trend than ones in Fig. \ref{fig:ee_K_theta1} for correlated users (\emph{i.e.,} for $\theta = 1$). Indeed, we remark that the Stackelberg perform almost the same as the Nash game for $\theta = 0$, whereas, for $\theta = 1$, the gap between the Nash game and the Stackelberg game increases. More specifically, the Stackelberg model achieves an energy efficiency gain up to 25\% with respect to the Nash model for $K=4$ carriers.
As the number of carriers $K$ goes large, both configurations tend towards having the same average energy efficiency. This can be justified by the fact that, when the number of carriers increases, the probability that users transmit on different carriers is high (see Section \ref{sec:coord}) and thus, users are less sensitive to their degree of hierarchy in the system (see Prop. \ref{prop:soc_welf}). Interestingly, in both the independent and correlated users' cases, the Stackelberg game achieves almost the same energy efficiency as at the social welfare, which tends to validate results in Prop. \ref{prop:soc_welf}.

\begin{figure}[t]
\centering
\hspace*{-0.2cm}
\includegraphics[height = 4.4cm,width=8.5cm]{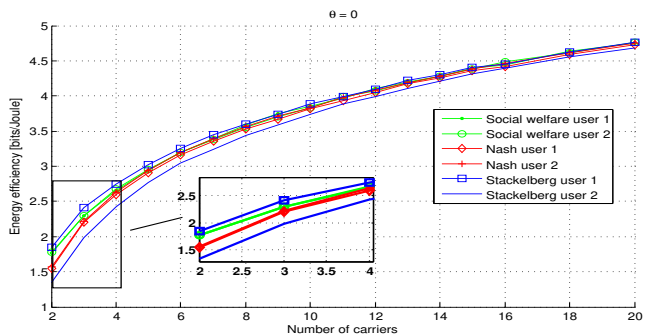}
%\vspace*{-6.1cm}
\caption{Per-user energy efficiency with independent users. User $1$ and user $2$ in the Stackelberg game refer to the leader and the follower respectively.}
\label{fig:ee_K_user}
\end{figure}

Fig. \ref{fig:ee_K_user} illustrates the per-user energy efficiency with independent users. Interestingly, we see from Fig. \ref{fig:ee_K_user} that, at the Stackelberg equilibrium, the energy efficiency of the follower in the Stackelberg game is smaller than in the simultaneous-move game. This suggests that, for the vast majority of cases, Situation 3) in Prop. \ref{prop:payoffs} is more likely to occur for a low number of carriers $K$. As $K$ increases, Situation 1) in Prop. \ref{prop:payoffs} is more likely to occur yielding the same energy efficiency for both the leader and the follower in the Stackelberg game as in the simultaneous-move game. This is justified by the fact that, with probability $1/K$, resp. $(K-1)/K$, users have the same, resp. different, best channels. It is then easy to see that, for low $K$, users are more likely to have the same best channels and interference is an issue in this case yielding to Situation 3) in Prop. \ref{prop:payoffs}, whereas, for sufficiently large $K$, users are more likely to have different best channels yielding to Situation 1) in Prop. \ref{prop:payoffs}. Moreover, Fig. \ref{fig:ee_K_user} also shows that it is profitable to be the leader which corresponds to what Prop. \ref{prop:lead_vs_follow} points out.

\begin{figure}[t]
\centering
\hspace*{-0.5cm}
\includegraphics[height = 4.4cm,width=9.5cm]{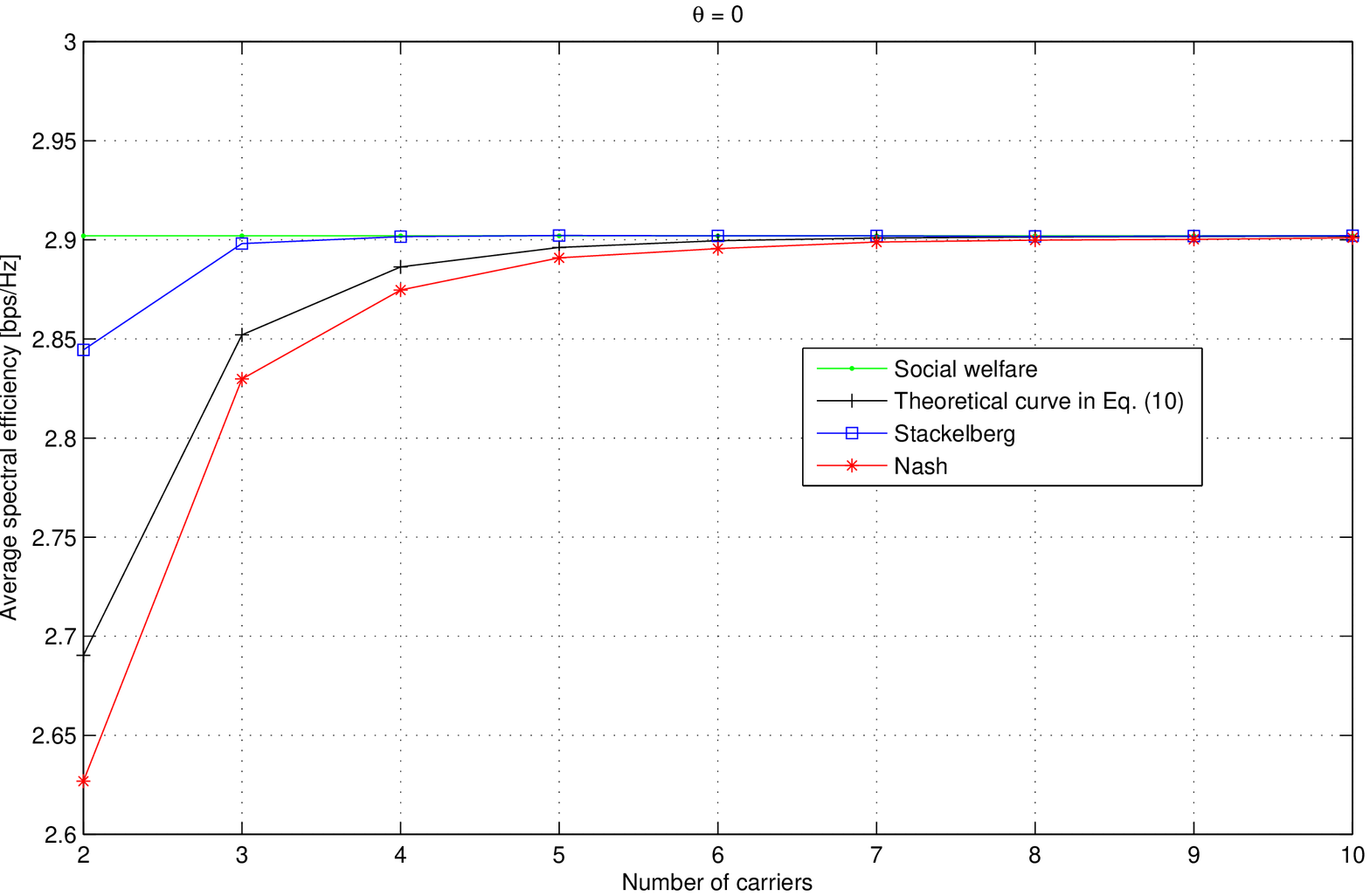}
%\vspace*{-6.1cm}
\caption{Average spectral efficiency with independent users.}
\label{fig:se-theta0}
%\end{figure}
%\begin{figure}[t]
\centering
\vspace*{0.5cm}
\hspace*{-0.5cm}
\includegraphics[height = 4.4cm,width=9.5cm]{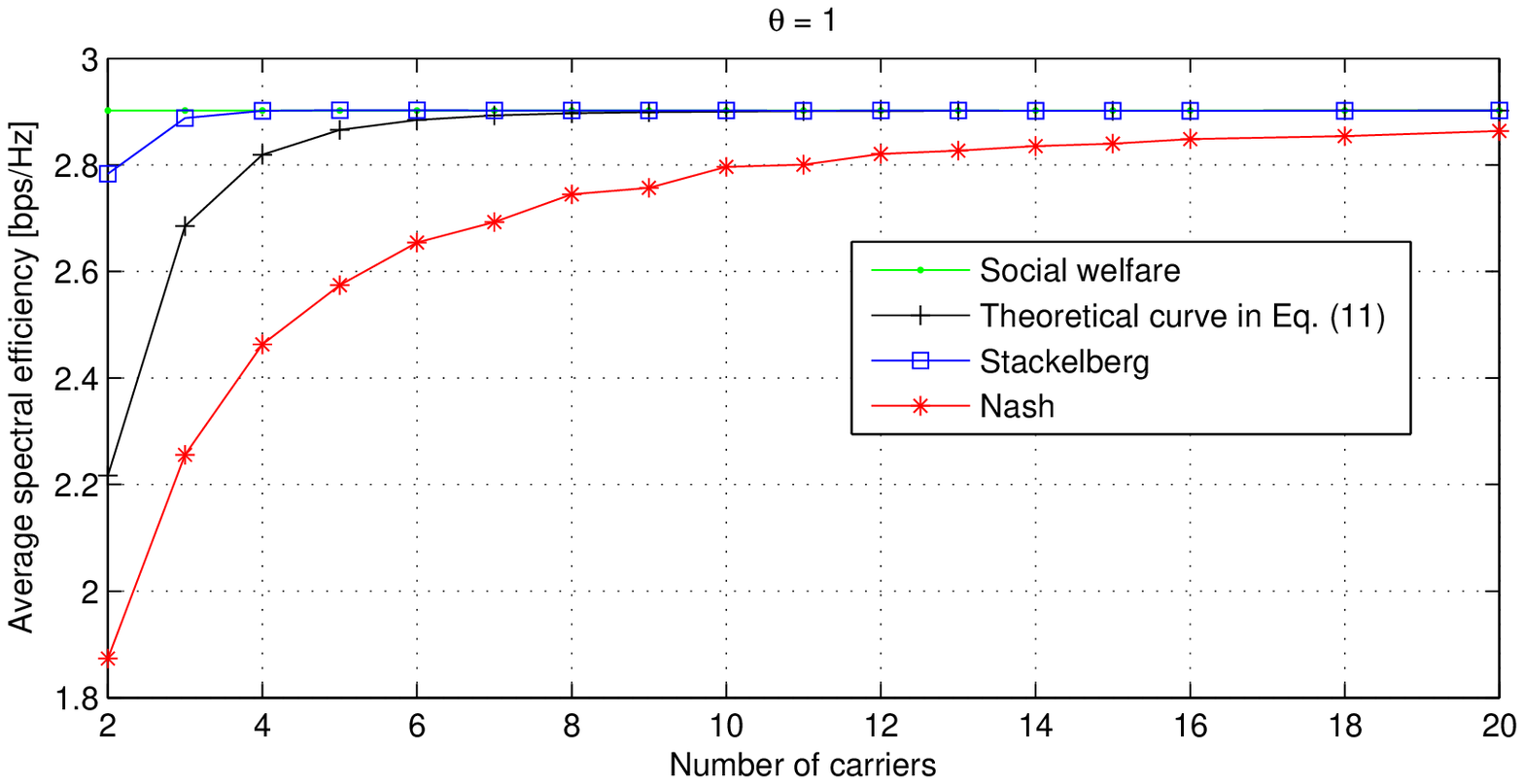}
%\vspace*{-5.8cm}
\caption{Average spectral efficiency with correlated users.}
\label{fig:se-theta1}
\end{figure}

\vspace*{-0.2cm}
\subsection{Spectral efficiency}

In Fig. \ref{fig:se-theta0} and \ref{fig:se-theta1}, we compare the closed-form expressions of the spectral efficiency derived in Eq. (\ref{eq:se}) for i.i.d. users (\emph{i.e.,} for $\theta = 0$) and in Eq. (\ref{eq:se1}) for correlated users (\emph{i.e.,} for $\theta = 1$) with the simulated spectral efficiency. Of
particular interest is the fact that the closed-form expressions turn out to be very tight. We can
also observe that the Stackelberg game performs better than the Nash game in terms of average spectral efficiency particularly for correlated users while still performing very close to the social welfare. As an example, for $K=2$ carriers, the Stackelberg game yields only a negligible spectral efficiency loss $0.05$ bps/Hz with respect to the social welfare and approximately $0.22$ bps/Hz of spectral efficiency gain beyond the Nash game.

\begin{figure}[t]
\centering
\hspace*{-0.5cm}
\includegraphics[height = 4.4cm,width=9.5cm]{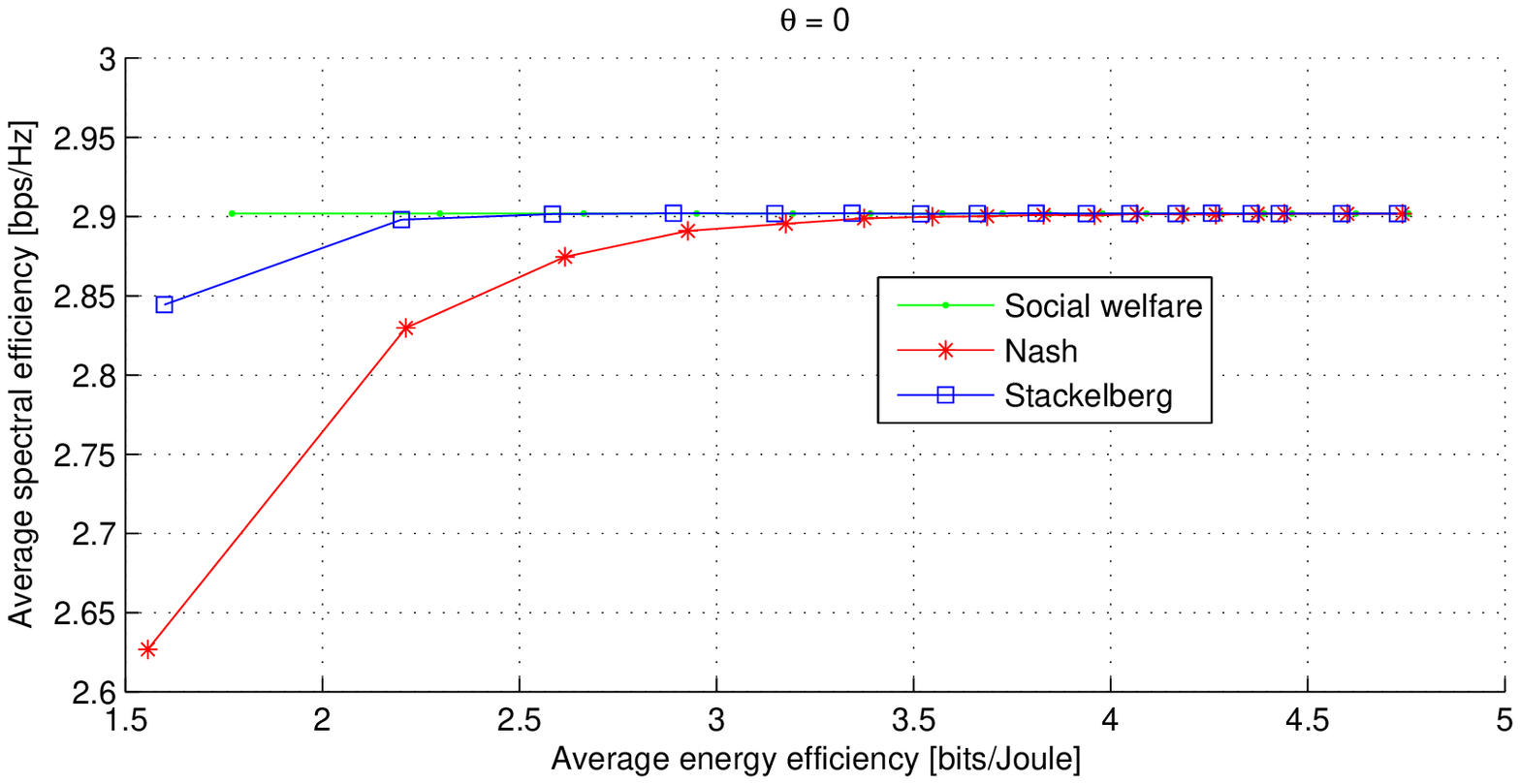}
%\vspace*{-5.6cm}
\caption{Spectral efficiency vs. Energy efficiency with independent users.}
\label{fig:se_ee-theta0}
%\end{figure}
%\begin{figure}[t]
\centering
\vspace*{0.5cm}
\hspace*{-0.5cm}
\includegraphics[height = 4.4cm,width=9.5cm]{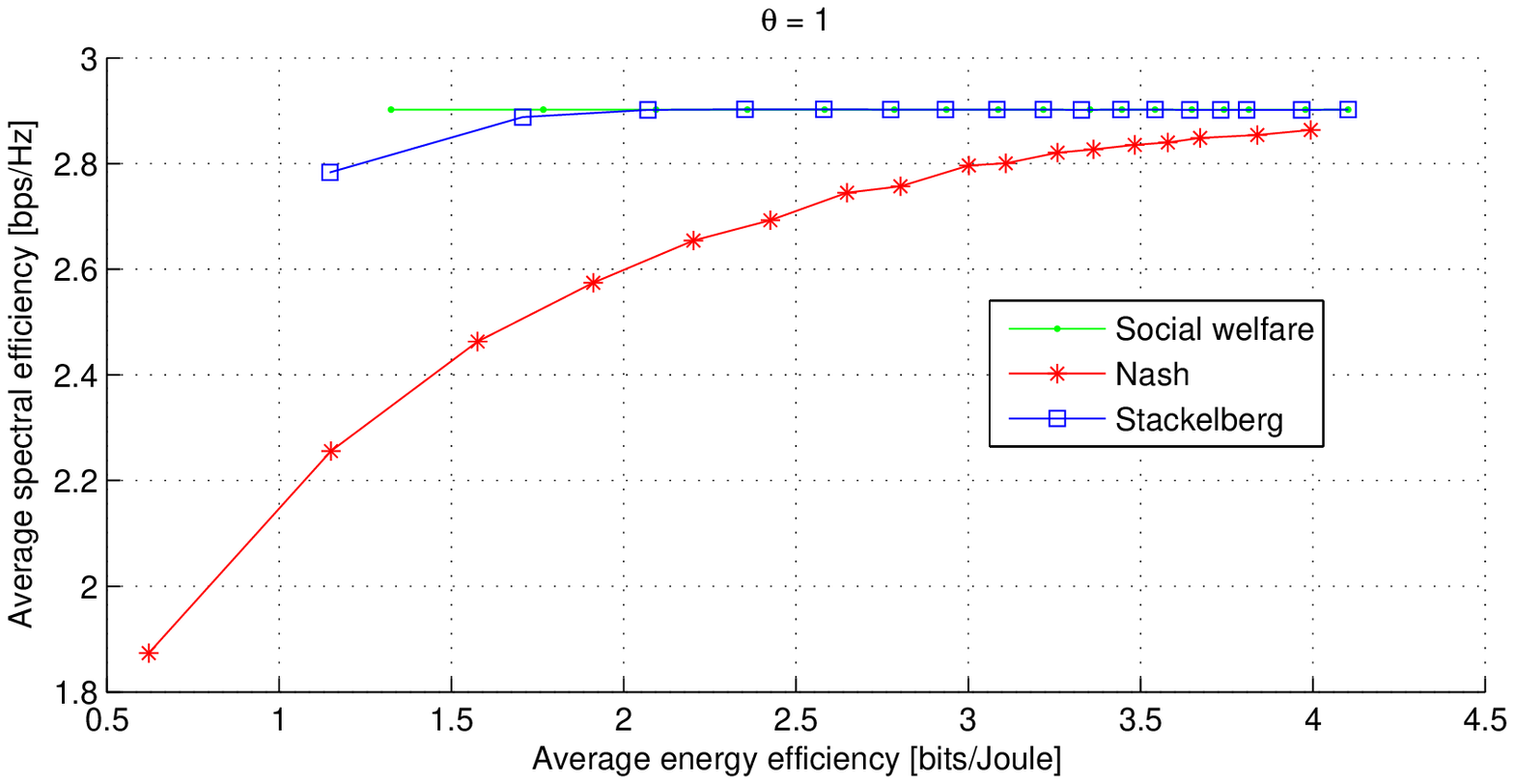}
%\vspace*{-5.6cm}
\caption{Spectral efficiency vs. Energy efficiency with correlated users.}
\label{fig:se_ee-theta1}
\end{figure}

\subsection{Spectral efficiency -- Energy efficiency Tradeoff}

In order to illustrate the balance between the achievable rate and energy consumption of the system, we plot in Fig.~\ref{fig:se_ee-theta0} and \ref{fig:se_ee-theta1} the spectral efficiency as a function of the energy efficiency for independent and correlated users respectively. Surprisingly, it is clearly shown that, for both the independent and correlated cases, the proposed Stackelberg decision approach achieves a flexible and desirable tradeoff between energy efficiency and throughput maximization compared to the social welfare and the Nash model. In particular, it is shown that the Stackelberg scheme maximizes the energy efficiency while still optimizing the spectral efficiency at the Stackelberg equilibrium. Notice that this contrasts with most related works so far in which the optimal
energy efficiency performance often leads to low spectral efficiency performance and vice versa \cite{tradeoff_green_commag11,DengRCZZL13,wcnc/AminMEH12,JoungESTJSAC2014}.
This feature has a great impact on the network performance and provides a convincing argument that hierarchical communication is the proper context to design and optimize energy efficient wireless networks.

\vspace{-0.2cm}
\section{Conclusion}\label{sec:conc}

The growing interest in energy efficient research from signal processing and communication communities has spurred an increasing interest in the recent years. There have been a large number of proposals for all communication layers, but the system infrastructure has not been clearly defined. In this paper, we have proposed a hierarchical game to model distributed joint power and channel allocation for multi-carrier energy efficient systems since it has the advantage of leading towards more realistic or even simpler distributed power control algorithms. We have established the existence of the Stackelberg equilibrium and gave its formal expression. The proposed scheme achieves better performances as compared to those of other existing schemes, notably the Nash model proposed in \cite{meshkati-jsac-2006}. In particular, we have proved that introducing hierarchy across users induces a spectrum orthogonalization which substantially improves system performances. For the first time, we have derived the spectral efficiency of such a model with exact expressions for the throughput scaling. The proposed scheme can achieve a spectral efficiency scaling of $\log_2(1+\gamma^*)\left[1-\mathcal{O}(K^{-(1+\gamma^*)})\right]$, while a vanishing fraction of the carriers may suffer from mutual interference as the number of the carriers goes large. Simulation results have been presented to exhibit the effectiveness of the proposed scheme to balance the achievable rate and energy consumption of the system.

%\section{Acknowledgments}
%This work was partially supported by Wroclaw University of Technology under grant number S40043/K1101.

\bibliographystyle{IEEEtran}
\bibliography{mybib}

\appendix

\subsection{Proof of Proposition \ref{multi:StackNash}}
\begin{lemma}
\label{lemma:Tardos}
For any finite sequence of $n$ pairs $(a_k,b_k)$ such that $a_k\geq 0$ and $b_k>0$ the following inequality is true:
$$\frac{\sum_{k=1}^K a_k}{\sum_{k=1}^K b_k}\leq \max\{\frac{a_k}{b_k},k=1,\ldots,n\}.$$
The equality is only possible if each ratio $\frac{a_k}{b_k}$ is equal.
\end{lemma}
\begin{IEEEproof}
We proceed by induction with respect to $n$. For $n=2$, let us assume that the hypothesis is not true and thus:
$$\frac{a_1+a_2}{b_1+b_2}>\frac{a_1}{b_1}\quad\mbox{and}\quad\frac{a_1+a_2}{b_1+b_2}>\frac{a_2}{b_2}.$$
This can be rewritten as
$$a_1b_1+a_2b_1>a_1b_1+a_1b_2\quad\mbox{and}\quad a_1b_2+a_2b_2>a_2b_1+a_2b_2$$
or equivalently $a_2b_1>a_1b_2>a_2b_1$, which is a contradiction.

{\normalsize Next, assume that our hypothesis is true for any $l<K$.
Then, we can proceed as follows:}

{\scriptsize\begin{eqnarray*}
\frac{\sum_{k=1}^K a_k}{\sum_{k=1}^K b_k}\leq\frac{\sum_{k=1}^{K-1} a_k+a_K}{\sum_{k=1}^{K-1} b_k+b_K}&\leq&\max\left\{\frac{\sum_{k=1}^K a_k}{\sum_{k=1}^K b_k},\frac{a_K}{b_K}\right\}\\
&\leq& \max\left\{\max\{\frac{a_k}{b_k},k=1,\ldots,K-1\},\frac{a_K}{b_K}\right\}\\
&=&\max\{\frac{a_k}{b_k},k=1,\ldots,K\}
\end{eqnarray*}}
If there is at least one pair $(a_k,b_k)$, whose ratio is bigger than the other ones we can show along the same lines that the inequality is strong (we only need to take these $a_k$ and $b_k$ from the sums $\sum_{k=1}^K a_k$, $\sum_{k=1}^K b_k$ in the above considerations instead of $a_K$ and $b_K$.\\
\end{IEEEproof}

Now we can prove Proposition \ref{multi:StackNash}.\\
\begin{IEEEproof}
Note that by Lemma \ref{lemma:Tardos}
$$u_1(p_1,p_2)=\frac{\sum_{k=1}^KR_1f(\gamma_1^k)}{\sum_{k=1}^Kp_1^k}\leq\max_k\frac{R_1f(\gamma_1^k)}{p_1^k},$$
so the leader in the Stackelberg game cannot use more than one carrier simultaneously, as decreasing power to zero on every carrier different from the one realizing maximum above would be beneficial. Thus he will choose only one carrier for which
$$\hat{f}_1^k(p_1^k)=\frac{f(\gamma_1^k)}{p_1^k}=\frac{1}{p_1^k}f(\frac{g_1^kp_1^k}{\sigma^2+g_2^k\overline{p}_2^k(p_1)})$$
(where $\overline{p}_2^k$ is computed according to (\ref{eq:follower})) is the greatest. Note however that since the follower will chose only one carrier, $\hat{f}_1^k(p_1^k)$ will be equal to $\frac{f(\frac{g_1^kp_1^k}{\sigma^2(1+\gamma^*)+\gamma^*g_1^kp_1^k})}{p_1^k}$ only for one carrier, say carrier $k^*$, and for any other carrier it will be equal to $\frac{f(\frac{g_1^kp_1^k}{\sigma^2})}{p_1^k}$, which is maximized for $p_1^k=\frac{\gamma^*\sigma^2}{g_1^k}$ and then equal to $\frac{f(\gamma^*)g_1^k}{\gamma^*\sigma^2}$. But this last value depends on the carrier only through $g_1^k$, so will be maximized for $k=B_1$ if only $k^*\neq B_1$. Thus the equilibrium strategy of the leader will put all the power on carrier $B_1$ in that case. If $k^*=B_1$, then the biggest value of $\hat{f}_1^k(p_1^k)$ for $k\neq k^*$ will be for $k=S_1$, and either all the power of the leader will be put on this carrier or on $k^*=B_1$.

As for the follower, by Proposition \ref{prop:follower-power} his best response is always to put all his power on the carrier maximizing
$\hat{h}_2^k(p_1^k)=\frac{g_2^k}{\sigma^2+g_1^kp_1^k}$, which will be equal to $\frac{g_2^k}{\sigma^2}$ for all but one carrier. Now the reasoning made for the leader can be applied here as well.\\
\end{IEEEproof}

\subsection{Proof of Proposition \ref{multi:power_alloc}}
\begin{IEEEproof}
First consider the case when $B_1\neq B_2$. The biggest possible value of the ratio $\frac{f(\gamma_n^k)}{p_n^k}$ obtainable for player $n$ on a single carrier (when his opponent does not maximize his payoff, but also the payoff of player $n$) is $\frac{f(\gamma^*)R_n}{\gamma^*\sigma^2}\max_kg_n^k=\frac{f(\gamma^*)R_n}{\gamma^*\sigma^2}g_n^{B_n}$. Just this is obtained by both players when they apply strategies $\overline{p}_n$ defined in the theorem. Thus none of them will be interested in changing his strategy.\\

Now we move to the case when $B_1=B_2$. Suppose the leader uses only carrier $B_1$ in his equilibrium strategy. Then, by Proposition \ref{multi:StackNash} the follower uses one of carriers $B_1=B_2$ or $S_2$. If he uses $B_1$ then by Proposition \ref{prop:follower-power} the following has to be true:
$$\hspace*{0.3cm}\hat{h}_2^{B_1}(p_1^{B_1})=\frac{g_2^{B_1}}{\sigma^2+g_1^{B_1}p_1^{B_1}}\geq \frac{g_2^{S_2}}{\sigma^2}=\hat{h}_2^{S_2}(p_1^{S_2}).$$
Rewriting this we obtain that the follower chooses $B_1$ when
\vspace{.5cm}\begin{equation}
\label{case1_cond}
p_1^{B_1}\leq \frac{\sigma^2(g_2^{B_1}-g_2^{S_2})}{g_1^{B_1}g_2^{S_2}}
\end{equation}
%\vspace{-0.5cm}
\\and $S_2$ otherwise. Having this in mind, we can compute the utility of the leader at the equilibrium using carrier $B_1$, namely
\begin{equation}
\label{leader_utility1}
R_1\frac{f(\frac{g_1^{B_1}p_1^{B_1}}{\sigma^2(1+\gamma^*)+\gamma^*g_1^{B_1}p_1^{B_1}})}{p_1^{B_1}}
\end{equation} when $p_1^{B_1}\leq\frac{\sigma^2(g_2^{B_1}-g_2^{S_2})}{g_1^{B_1}g_2^{S_2}}$ and
\begin{equation}
\label{leader_utility2}
R_1\frac{f(\frac{g_1^{B_1}p_1^{B_1}}{\sigma^2})}{p_1^{B_1}}
\end{equation}
otherwise. Next we need to find the values of $p_1^{B_1}$ maximizing (\ref{leader_utility1}) and (\ref{leader_utility2}) respectively.
Before we obtain the first one we rewrite the SINR in that case in the following way:
\begin{equation}
\label{leader_sinr1}
\gamma_1^{B_1}=\frac{1}{\gamma^*}\left( 1-\frac{1}{1+\frac{\gamma^*g_1^{B_1}p_1^{B_1}}{\sigma^2(1+\gamma^*)}}\right)
\end{equation}
and differentiate it with respect to $p_1^{B_1}$, obtaining:
\begin{eqnarray}
\label{sinr_diff1}
\frac{\partial \gamma_1^{B_1}}{\partial p_1^{B_1}}&=&\frac{g_1^{B_1}}{\sigma^2(1+\gamma^*)(1+\frac{\gamma^*}{\sigma^2(1+\gamma^*)}g_1^{B_1}p_1^{B_1})^2}\\\nonumber
&=&\frac{g_1^{B_1}\sigma^2(1+\gamma^*)}{{(\sigma^2(1+\gamma^*)+\gamma^*g_1^{B_1}p_1^{B_1})^2}} \\\nonumber
&=&\frac{1}{p_1^{B_1}}\frac{\sigma^2(1+\gamma^*)}{g_1^{B_1}p_1^{B_1}\gamma^*}(\gamma_1^{B_1})^2\gamma^*
\end{eqnarray}
Next, we can transform (\ref{leader_sinr1}) into
\begin{equation}
\label{sinr_transformed}
\frac{\gamma^*g_1^{B_1}p_1^{B_1}}{\sigma^2(1+\gamma^*)}=\frac{\gamma^*\gamma_1^{B_1}}{1-\gamma^*\gamma_1^{B_1}}.
\end{equation}
and put it into (\ref{sinr_diff1}), obtaining:
\begin{equation}
\label{sinr_diff2}
\frac{\partial \gamma_1^{B_1}}{\partial p_1^{B_1}}=\frac{1}{p_1^{B_1}}\frac{1-\gamma^*\gamma_1^{B_1}}{\gamma^*\gamma_1^{B_1}}(\gamma_1^{B_1})^2\gamma^*=
\frac{1}{p_1^{B_1}}(\gamma_1^{B_1}-\gamma^*(\gamma_1^{B_1})^2).
\end{equation}
Now we write the first order condition for the maximization of (\ref{leader_utility1}):
$$0=\frac{\partial(\frac{R_1f(\gamma_1^{B_1})}{p_1^{B_1}})}{\partial p_1^{B_1}}=R_1\frac{-f(\gamma_1^{B_1})+f^{\prime}(\gamma_1^{B_1})\frac{\partial \gamma_1^{B_1}}{\partial p_1^{B_1}}p_1^{B_1}}{(p_1^{B_1})^2}.$$
If we substitute (\ref{sinr_diff2}) into it, we obtain the following equation:
\begin{equation}
\label{utility_max1}
-f(\gamma_1^{B_1})+(\gamma_1^{B_1}-\gamma^*(\gamma_1^{B_1})^2)f^{\prime}(\gamma_1^{B_1}).
\end{equation}
If we find the best solution to this equation (that is, maximizing $\frac{f(\gamma_1^{B_1})}{p_1^{B_1}}$), $\beta^{*}$, we get the power allocation of the leader in case (\ref{case1_cond}), which can be computed from (\ref{sinr_transformed}) as
\begin{equation}
\label{alloc1}
p^{**}=\frac{\beta^{*}\sigma^2(1+\gamma^*)}{g_1^{B_1}(1-\gamma^*\beta^{*})}.
\end{equation}

Similarly, when we write the first order condition for the maximization of (\ref{leader_utility2}), we obtain
$$0=\frac{\partial(\frac{R_1f(\gamma_1^{B_1})}{p_1^{B_1}})}{\partial p_1^{B_1}}=\frac{-f(\gamma_1^{B_1})+\gamma_1^{B_1}f^{\prime}(\gamma_1^{B_1})}{(p_1^{B_1})^2},$$
whose unique solution is $\gamma^*$. The corresponding value of $p_1^{B_1}$ is
\begin{equation}
\label{alloc2}
p^*=\frac{\gamma^*\sigma^2}{g_1^{B_1}}.
\end{equation}
Now, we put (\ref{alloc1}) and (\ref{alloc2}) in (\ref{leader_utility1}) and (\ref{leader_utility2}) respectively, obtaining the value functions corresponding to $p^*$ and $p^{**}$:

{\scriptsize
\beqq
V_{B_1}=\frac{f(\beta^{*})(1-\gamma^*\beta^{*})g^{B_1}_1R_1}{\beta^{*}\sigma^2(1+\gamma^*)}
\quad\text{and}\quad U_{B_1}=\frac{f(\gamma^*)g^{B_1}_1R_1}{\gamma^*\sigma^2}
\eeqq}
Note that the first one is always smaller than the second one (because $\gamma^*$ maximizes the ratio $\frac{f(x)}{x}$ and $\gamma^*,\beta^{*}>0$). So, in case $p^*$ satisfies the condition opposite to (\ref{case1_cond}), the leader will choose to transmit on $B_1$ with this power, while the follower will choose (according to (\ref{prop:follower-power})) to transmit on $S_2$ with power $\frac{\gamma^*\sigma^2}{g_2^{S_2}}$.

Next, when $p^*$ satisfies (\ref{case1_cond}), the situation becomes more complex. The leader has to choose between one of the three possibilities: to choose the power $p^{**}$ on carrier $B_1$, giving him the value of $V_{B_1}$, to choose power
$\hat{p}=\frac{\hat{\gamma}\sigma^2}{g_1^{B_1}}$
on carrier $B_1$, which now gives the biggest value in case the follower chooses to use carrier $S_2$,
$W_{B_1}=\frac{f(\hat{\gamma})g_1^{B_1}R_1}{\hat{\gamma}\sigma^2}$,
or to choose to use his second-best carrier $S_1$ instead of $B_1$, with power
$p_{S_1}=\frac{\gamma^*\sigma^2}{g_1^{S_1}},$
which would give him the value
$U_{S_1}=\frac{f(\gamma^*)g^{S_1}_1R_1}{\gamma^*\sigma^2}.$
Choosing the biggest one from $V_{B_1}$, $W_{B_1}$ and $U_{S_1}$ will give the leader's equilibrium payoff (and corresponding equilibrium strategy) in the Stackelberg game, unless $V_{B_1}$ is not the biggest value obtainable by the leader in case (\ref{case1_cond}). This is only possible when the biggest value of (\ref{leader_utility1}) is obtained on one of the ends of the interval $(0,\frac{\hat{\gamma}}{1+\gamma^*(1+\hat{\gamma})}]$. Thus, we compute these two values:
$$V^0_{B_1}=\lim_{\gamma\rightarrow 0}\frac{R_1f(\gamma)(1-\gamma^*\gamma)g_1^{B_1}}{\gamma\sigma^2(1+\gamma^*)}=\frac{R_1f^{\prime}(0)g_1^{B_1}}{\sigma^2(1+\gamma^2)},$$
\begin{eqnarray*}
V^1_{B_1}&=&\frac{R_1f(\frac{\hat{\gamma}}{1+\gamma^*(1+\hat{\gamma})})(1-\gamma^*\frac{\hat{\gamma}}{1+\gamma^*(1+\hat{\gamma})})g_1^{B_1}}{\frac{\hat{\gamma}}{1+\gamma^*(1+\hat{\gamma})}\sigma^2(1+\gamma^*)}\\\nonumber
&=&R_1\frac{f(\frac{\hat{\gamma}}{1+\gamma^*(1+\hat{\gamma})})g_1^{B_1}}{\hat{\gamma}\sigma^2}
\end{eqnarray*}
$V^1_{B_1}$ is clearly smaller than $W_{B_1}$, so it cannot be the biggest value obtained by the leader. The value $V^0_{B_1}$ though can be the biggest one, and so in case $V^0_{B_1}$ is bigger than $\max\{ V_{B_1},W_{B_1},U_{S_1}\}$ it is optimal for the leader to use the smallest power possible on carrier $B_1$ (which is not an equilibrium strategy, as for any arbitrarily small power there exists a smaller power, for which the value function of the leader is closer to $V^0_{B_1}$. The power allocations of the follower in each of the cases of (\ref{case1_cond}) are computed according to Proposition \ref{prop:follower-power}.\\
\end{IEEEproof}

\subsection{Proof of Proposition \ref{multi:eps_equilibria}}
\begin{IEEEproof}
The inequality (\ref{eps_cond}) is a rewriting of the condition $V^0_{B_1}>\max\{ V_{B_1},W_{B_1},U_{S_1}\}$, appearing in the proof of Theorem \ref{multi:StackNash}, where the optimal behavior of the leader in the case when this condition, together with $B_1=B_2$ and $\hat{\gamma}>\gamma^*$ is satisfied, was also described. The behavior of the follower follows from Proposition \ref{prop:follower-power}.
\end{IEEEproof}

\subsection{Proof of Proposition \ref{multi:no_coordination_region}}
\begin{IEEEproof}
By Proposition \ref{multi:power_alloc}, no orthogonalization between the players is only possible if $B_1=B_2$,
\begin{equation}
\hat{\gamma}>\gamma^*
\label{gamma_nneq}
\end{equation}
and
\begin{equation}
\label{vw_nneq}
\max\{ V_{B_1}^0,V_{B_1}\}>\max\{W_{B_1},U_{S_1}\}.
\end{equation}
(\ref{gamma_nneq}) can be rewritten as
$\frac{g_2^{B_2}-g_2^{S_2}}{g_2^{S_2}}>\gamma^*$,
which is then equivalent to
$g_2^{B_2}>(1+\gamma^*)g_2^{S_2}.$
On the other hand (\ref{vw_nneq}) implies that $V_{B_1}^0>U_{S_1}$, which can be written as
$\frac{f^{\prime}(0)g_1^{B_1}}{1+\gamma^*}>\frac{f(\gamma^*)g_1^{S_1}}{\gamma^*}.$
Now, using the definition of $\gamma^*$ and the fact that $f^{\prime}(0)<f^{\prime}(\gamma^*)$ (see \cite{rodriguez-globecom-2003}) we can conclude that
$\frac{f^{\prime}(\gamma^*)g_1^{B_1}}{1+\gamma^*}>f^{\prime}(\gamma^*)g_1^{S_1}$,
which implies
$g_1^{B_1}>(1+\gamma^*)g_1^{S_1},$
ending the proof.\\
\end{IEEEproof}

\subsection{Proof of Proposition \ref{multi:no_coordination_prob}}
\begin{IEEEproof}
{\normalsize
By Proposition \ref{multi:no_coordination_region}, no orthogonalization is only possible if $B_1=B_2$ and $g_n^{B_n}\geq(1+\gamma^*)g_n^{S_n}$ for $n=1,2$ (which is an exact condition for no orthogonalization in the simultaneous-move model). The probability of this can be computed as}

{\scriptsize\begin{equation}
\label{no_coord_prob0}
1-\left[ \frac{K-1}{K}+\frac{\mathcal{P}}{K}+\frac{1-\mathcal{P}}{K}\frac{\mathcal{P}}{K}\right]=\frac{1-\mathcal{P}}{K}\left( 1-\frac{\mathcal{P}}{K}\right).
\end{equation}}
where $\mathcal{P}$ denotes the probability that for one of the players $g_i^{B_i}<(1+\gamma^*)g_i^{S_i}$. We can easily compute that

{\scriptsize$$\hspace*{-0.3cm}\mathcal{P}=K!\int_0^\infty dg_i^K\int_{g_i^K}^\infty dg_i^{K-1}\ldots\int_{g_i^3}^\infty dg_i^2\int_{g_i^2}^{(1+\gamma^*)g_i^2}\lambda^Ke^{-\lambda\sum_{k=1}^Kg_i^k}dg_i^1.$$}
If we introduce new variables $x^1=\lambda(g_i^1-g_i^2)$, $x^2=\lambda(g_i^2-g_i^3)$, $\ldots$, $x^{K-1}=\lambda(g_i^{K-1}-g_i^k)$, $x^K=\lambda g_i^K$, we can write it as
{\small$$\hspace*{-0.3cm}K!\int_0^\infty dx^K\int_0^\infty dx^{K-1}\ldots\int_0^\infty dx^2\int_0^{\gamma^*\sum_{k=2}^Kx^k}e^{-\sum_{k=1}^Kkx^k}dx^1$$}
and further as
$1-\frac{K!}{(2+\gamma^*)\ldots(K+\gamma^*)}$.
If we substitute it into the bound of no orthogonalization probability (\ref{no_coord_prob0}), we obtain

{\scriptsize\begin{eqnarray*}
&\ds\frac{(K-1)!}{(2+\gamma^*)\ldots(K+\gamma^*)}\left(\frac{K-1}{K}+\frac{(K-1)!}{(2+\gamma^*)\ldots(K+\gamma^*)}\right)
=\ds(1+\gamma^*)\mathcal{B}(1+\gamma^*,K)\left[ \frac{K-1}{K}+(1+\gamma^*)\mathcal{B}(1+\gamma^*,K)\right].
\end{eqnarray*}}
It can be immediately seen that this is no less than $(1+\gamma^*)\mathcal{B}(1+\gamma^*,K)$.
The fact that this last quantity is $\mathcal{O}(K^{-(1+\gamma^*)})$ is well known (see \emph{e.g.,} pp. 263 in \cite{Abramowitz72}).
\end{IEEEproof}

\subsection{Proof of Proposition \ref{prop:payoffs}}
\begin{IEEEproof}
{\normalsize
First note that the players in the Stackelberg game both use carrier $B_1=B_2$ in ($\epsilon$-)equilibrium when $\frac{g_1^{B_1}}{g_1^{S_1}}$ and $\frac{g_2^{B_2}}{g_2^{S_2}}$ satisfy}

{\scriptsize\begin{equation}
\label{eq:gammatt}
\max\left\{ \frac{f(\beta^{*})(1-\gamma^*\beta^{*})}{\beta^{*}(1+\gamma^*)},\frac{f^{\prime}(0)}{1+\gamma^*}\right\}\frac{g_1^{B_1}}{g_1^{S_1}}>\frac{f(\gamma^*)}{\gamma^*},
\end{equation}
\begin{equation}
\label{eq:gammat}
\max\left\{ \frac{f(\beta^{*})(1-\gamma^*\beta^{*})}{\beta^{*}(1+\gamma^*)},\frac{f^{\prime}(0)}{1+\gamma^*}\right\}>\frac{f(\frac{g_2^{B_2}}{g_2^{S_2}}-1)}{\frac{g_2^{B_2}}{g_2^{S_2}}-1},
\end{equation}}
which is true for $\frac{g_1^{B_1}}{g_1^{S_1}}$ and $\frac{g_2^{B_2}}{g_2^{S_2}}$ big enough (where the latter is a consequence of the fact that the RHS of (\ref{eq:gammat}) converges to $0$ as $\frac{g_2^{B_2}}{g_2^{S_2}}$ goes to infinity). If we intersect the set obtained with the set where $\frac{g_1^{B_1}}{g_1^{S_1}}$ and $\frac{g_2^{B_2}}{g_2^{S_2}}$ are bigger than $\frac{1}{1-\gamma^*}$ we get the desired set where there is no orthogonalization in equilibria of both the Stackelberg and simultaneous-move games.

Now let us compute the payoffs of the follower in this situation. The payoff in the simultaneous-move game equals
$\frac{f(\gamma^*)g_2^{B_2}(1-\gamma^*)R_2}{\gamma^*\sigma^2}$,
while that in the Stackelberg game is
$\frac{f(\gamma^*)g_2^{B_2}(1-\gamma^*\beta^{*})R_2}{\gamma^*\sigma^2(1+\beta^{*})}$.
The latter is bigger if
$1-\gamma^*<\frac{1-\gamma^*\beta^{*}}{1+\beta^{*}}$,
which is equivalent to $\gamma^*>\beta^{*}$. This is always true, as any solution to (\ref{gamma**}) has to be smaller than $\gamma^*$.

Next, suppose that $\frac{g_1^{B_1}}{g_1^{S_1}}<1+\gamma^*$ and $\frac{g_2^{B_2}}{g_2^{S_2}}>1+\gamma^*$, and thus player 1 uses carrier $S_1$, while player 2 uses carrier $B_2$ in the only equilibrium of the simultaneous-move game. Then, to obtain the situation where it is player 1 who uses his best carrier and player 2 who uses his second-best one in the Stackelberg game, the inequality
$\frac{f(\frac{g_2^{B_2}}{g_2^{S_2}}-1)}{\frac{g_2^{B_2}}{g_2^{S_2}}-1}\frac{g_1^{B_1}}{g_1^{S_1}}>\frac{f(\gamma^*)}{\gamma^*}$ has to be true.
If we denote by $y(x)$ the solution\footnote{It follows from the fact that $\frac{f(x)}{x}$ is decreasing for $x>1+\gamma^*$ that there is always only one such $y$.} of the equation $\frac{f(x-1)}{x-1}y(x)=\frac{f(\gamma^*)}{\gamma^*}$, we may rewrite the above three inequalities as
\begin{equation}
\label{eq:phigamma}
y(\frac{g_2^{B_2}}{g_2^{S_2}})<\frac{g_1^{B_1}}{g_1^{S_1}}<1+\gamma^*,\quad \frac{g_2^{B_2}}{g_2^{S_2}}>1+\gamma^*.
\end{equation}

Now note that since the function $f$ is sigmoidal, $\frac{f(x-1)}{x-1}$ strictly decreases on the set $x>1+\gamma^*$. Combining this with the fact that $\frac{f((1+\gamma^*)-1)}{(1+\gamma^*)-1}\frac{1}{1}=\frac{f(\gamma^*)}{\gamma^*}$, one can see that for $x>1+\gamma^*$ the curve $y(x)$ is strictly increasing, and thus the set of pairs $(\frac{g_1^{B_1}}{g_1^{S_1}},\frac{g_2^{B_2}}{g_2^{S_2}})$ satisfying (\ref{eq:phigamma}) is not empty.

The payoffs of the follower in the simultaneous-move game and Stackelberg game (respectively) are in this situation
$\frac{f(\gamma^*)g_2^{B_2}R_2}{\gamma^*\sigma^2}$
and
$\frac{f(\gamma^*)g_2^{S_2}R_2}{\gamma^*\sigma^2}.$
Clearly the former is greater than the latter.

The final case is obvious, as in case when $B_1\neq B_2$ the strategies the players use are the same in the simultaneous-move and Stackelberg games.\\
\end{IEEEproof}

\subsection{Proof of Proposition \ref{prop:lead_vs_follow}}
\begin{IEEEproof}
To prove this prop., we only need to compare the utilities for player 1 when he is the leader and when he is the follower in each of the cases of Prop. \ref{multi:power_alloc}.\\
\end{IEEEproof}

\subsection{Proof of Proposition \ref{prop:soc_welf}}
\begin{IEEEproof}
The first part of the proposition is obvious. To prove 1) of the second part first note that the social welfare in equilibrium of the simultaneous-move game can only be bigger than that in Stackelberg equilibrium when the payoff of the follower in the Stackelberg game decreases. This is only possible when the carrier he uses in equilibrium changes from $B_1=B_2$ in the simultaneous-move game to $S_2$ in the Stackelberg game. In such a case his utility changes from $\frac{f(\gamma^*)g_2^{B_2}R_2}{\gamma^*\sigma^2}$ to $\frac{f(\gamma^*)g_2^{S_2}R_2}{\gamma^*\sigma^2}$ if the leader also changes the carrier he uses from $S_1$ to $B_1$ or from $\frac{f(\gamma^*)(1-\gamma^*)g_2^{B_2}R_2}{\gamma^*\sigma^2}$ to $\frac{f(\gamma^*)g_2^{S_2}R_2}{\gamma^*\sigma^2}$ if the leader uses carrier $B_1$ in both simultaneous-move and Stackelberg equilibria. On the other hand the utility of the leader in the Stackelberg equilibrium is $\frac{f(\gamma^*)g_1^{B_1}R_1}{\gamma^*\sigma^2}$ in the former case and not smaller than $\frac{f(\gamma^*)(1-\gamma^*)g_1^{B_1}R_1}{\gamma^*\sigma^2}$ in the latter one (this is because this is his utility in Nash equilibrium of the simultaneous game, and the utility of the leader increases in Stackelberg game). Straightforward computations yield the desired bound on the decrease of social welfare.

To prove part 2) first note that the maximum utility that can be obtained in this game is bounded above by
\begin{equation}
\frac{f(\gamma^*)g_1^{B_1}R_1}{\gamma^*\sigma^2}+\frac{f(\gamma^*)g_2^{B_2}R_2}{\gamma^*\sigma^2},
\label{sw_bound}
\end{equation}
as this is the sum of maximal utilities of both players (but not obtainable at the same time if $B_1=B_2$). Next note that if the leader uses carrier $S_1$ in Stackelberg equilibrium, the sum of the utilities of both players is $\frac{f(\gamma^*)g_1^{S_1}R_1}{\gamma^*\sigma^2}+\frac{f(\gamma^*)g_2^{B_2}R_2}{\gamma^*\sigma^2}$,
$\frac{R_1g_1^{B_1}+R_2g_2^{B_2}}{R_1g_1^{S_1}+R_2g_2^{B_2}}<\frac{R_1g_1^{B_1}+R_2g_2^{B_2}}{R_1g_1^{S_1}+R_2g_2^{S_2}}$ times less than (\ref{sw_bound}).
On the other hand if he uses $B_1$ in Stackelberg equilibrium, his utility cannot be smaller than $\frac{f(\gamma^*)g_1^{S_1}R_1}{\gamma^*\sigma^2}$, while that of the follower not less than $\frac{f(\gamma^*)g_2^{S_2}R_2}{\gamma^*\sigma^2}$ (if they were not, each of them would change his carrier to $S_1$ or $S_2$). But the sum of these utilities is $\frac{R_1g_1^{B_1}+R_2g_2^{B_2}}{R_1g_1^{S_1}+R_2g_2^{S_2}}$ times less than (\ref{sw_bound}).\\
\end{IEEEproof}

\subsection{Proof of Proposition \ref{prop:spec_ef}}
\begin{IEEEproof}
No orthogonalization in the simultaneous-move game is possible exactly when $B_1=B_2$ and $g_n^{B_n}\geq(1+\gamma^*)g_n^{S_n}$ for $n=1,2$. 1 minus the exact probability of that region is computed in Proposition \ref{multi:no_coordination_prob}, and this is also the lower bound on the same probability for the Stackelberg game. The spectral efficiency in case there is orthogonalization between the players can be computed as the expected value of $\log_2(1+\gamma)$ over this region. Note however that $\gamma\equiv\gamma^*$ there, and so the bound on spectral efficiency is exactly $\log_2(1+\gamma^*)$ times (the bound on) the probability of orthogonalization, which is $1-(1+\gamma^*)\mathcal{B}(1+\gamma^*,K)\left[ \frac{K-1}{K}+(1+\gamma^*)\mathcal{B}(1+\gamma^*,K)\right]$.
\end{IEEEproof}

\end{document}